\documentclass{aa}  

\usepackage{graphicx}
\usepackage{txfonts}
\usepackage{newtxtext,newtxmath}

\usepackage{graphicx}	% Including figure files
\usepackage{amsmath}	% Advanced maths commands
\usepackage{amssymb}	% Extra maths symbols
\usepackage{xcolor}
\usepackage[colorlinks = true,linkcolor = blue,citecolor = blue]{hyperref}
\usepackage{caption}
\usepackage{subcaption}
\newcommand{\lambdaint}{\lambda_{\textrm{int}}}
\newcommand{\kappaext}{\kappa_{\textrm{ext}}}
\newcommand{\DsDds}{D_{\textrm{s}}/D_{\textrm{ds}}}
\newcommand{\Dd}{D_{\textrm{d}}}
\newcommand{\Ds}{D_{\textrm{s}}}
\newcommand{\Dds}{D_{\textrm{ds}}}
\newcommand{\Ddt}{D_{\Delta\textrm{t}}}
\newcommand{\kmsmpc}{\rm km\,s^{-1}\,Mpc^{-1}}
\newcommand{\kms}{\rm km\,s^{-1}}

\newcommand{\sref}[1]{Section~\ref{#1}}
\newcommand{\fref}[1]{Figure~\ref{#1}}
\newcommand{\tref}[1]{Table~\ref{#1}}
\newcommand{\eref}[1]{Equation~(\ref{#1})}

\begin{document}

   \title{TDCOSMO VI: Distance Measurements in Time-delay Cosmography under the Mass-sheet transformation}

   %\subtitle{Breaking the MST in TDC in generic cosmologies}

   \author{Geoff~C.-F.~Chen
          ,\inst{1}\fnmsep\thanks{Current email address: gcfchen@astro.ucla.edu}
          Christopher~D.~Fassnacht,\inst{1}
          Sherry~H.~Suyu,\inst{2,3,4}
          Ak{\i}n~Y{\i}ld{\i}r{\i}m,\inst{2}
          Eiichiro Komatsu,\inst{2,5}\\\and
          Jos$\acute{\textrm{e}}$ Luis Bernal\inst{6}
          }
        \institute{Department of Physics and Astronomy, University of California, Davis, CA 95616, USA
             \and 
             Max Planck Institute for Astrophysics, Karl-Schwarzschild-Strasse 1, D-85740 Garching, Germany
             \and 
             Physik-Department, Technische Universit\"{a}t M\"{u}nchen, James-Franck-Stra{\ss}e 1, 85748 Garching, Germany 
             \and 
             Academia Sinica Institute of Astronomy and Astrophysics (ASIAA), 11F of ASMAB, No.1, Section 4, Roosevelt Road, Taipei, 10617, Taiwan
             \and
             Kavli IPMU (WPI), UTIAS, The University of Tokyo, Kashiwa, Chiba 277-8583, Japan
             \and
             Department of Physics and Astronomy, Johns Hopkins University, 3400 North Charles
             Street, Baltimore, Maryland 21218, USA
             }

   \date{Received xxx, xxxx; accepted xxx, xxxx}

% \abstract{}{}{}{}{} 
% 5 {} token are mandatory
 
  \abstract{Time-delay cosmography with gravitationally lensed quasars plays an important role in anchoring the absolute distance scale and hence measuring the Hubble constant, $H_{0}$, independent of traditional distance ladder methodology. A current potential limitation of time delay distance measurements is the ``mass-sheet transformation'' (MST) which leaves the lensed imaging unchanged but changes the distance measurements and the derived value of $H_0$. 
  In this work we show that the standard method of addressing the MST in time delay cosmography, through a combination of high-resolution imaging and the measurement of the stellar velocity dispersion of the lensing galaxy, depends on the assumption that the ratio, $\DsDds$, of angular diameter distances to the background quasar and between the lensing galaxy and the quasar can be constrained.  
  This is typically achieved through the assumption of a particular cosmological model. %such as $\Lambda$CDM. %and hence distance measurements are  cosmological-model-dependent quantities.
  %We use mock data to demonstrate that using cosmological model to break the internal MST can lead to bias on the distance measurement, while using the relative distance indicators such as, supernovae type Ia and baryon acoustic oscillations can yield accurate cosmological-model-independent distance measurements.
  Previous work (TDCOSMO IV) addressed the mass-sheet degeneracy and derived $H_{0}$ under the assumption of $\Lambda$CDM model. 
  In this paper we show that the mass sheet degeneracy can be broken without relying on a specific cosmological model by combining lensing with relative distance indicators such as supernovae type Ia and baryon acoustic oscillations, which constrain the shape of the expansion history and hence $\DsDds$. With this approach, we demonstrate that the mass-sheet degeneracy can be constrained in a cosmological-model-independent way, and hence model-independent distance measurements in time-delay cosmography under mass-sheet transformations can be obtained.
  }
  % context heading (optional)
  % {} leave it empty if necessary  
   
  % aims heading (mandatory)
   
  % methods heading (mandatory)
   
  % results heading (mandatory)
   
  % conclusions heading (optional), leave it empty if necessary 
   
\titlerunning{Distance measurements in TDC under MST}
\authorrunning{G. C.-F. Chen et al.}
   \maketitle
%
%-------------------------------------------------------------------

\section{Introduction}
The Hubble constant ($H_{0}$) is one of the most important parameters in cosmology. Its value directly sets the age, the size, and the critical density of the Universe. 
Despite the great success of the $\Lambda$CDM model \citep{KomatsuEtal11,HinshawEtal13,planck18parameter}, a stringent challenge to the model comes from a discrepancy between the extremely precise $H_{0}$ ($= 67.4\pm0.5~\kmsmpc$) value derived from Planck measurements of the cosmic microwave background (CMB) anisotropies under the assumption of $\Lambda$CDM \citep{planck18parameter}, and the $H_{0}$ value from direct measurements of the local Universe \citep{VerdeEtal19}. 

The recent direct $H_{0}$ measurements ($H_0 = 74.03\pm1.42~\kmsmpc$) from Type Ia supernovae (SN1a), calibrated by the traditional Cepheid distance ladder \cite[SH0ES collaboration;][]{RiessEtal19}, show a $4.4\sigma$ tension with the Planck results.  
However, a recent measurement of $H_{0} = 69.8\pm0.8(\textrm{stat})\pm1.7(\textrm{sys})~\kmsmpc$  from SN1a calibrated by the Tip of the Red Giant Branch (CCHP) agrees at the 1.2$\sigma$ level with Planck and at the 1.7$\sigma$ with SH0ES results \citep{FreedmanEtal19}.
The spread in these results, whether due to systematic effects \citep{Efstathiou20} or not, clearly demonstrates that it is crucial to test any single methodology by different and independent datasets.

Time-delay cosmography \citep[TDC; e.g.,][]{TreuMarshall16,SuyuEtal18} provides a technique to constrain $H_{0}$ at low redshift that is completely independent of the traditional distance ladder approach. 
When a quasar is strongly lensed by a galaxy, its multiple images have light-curves that are offset by a well-defined time delay, which depends on the mass profile of the lens and cosmological distances to the galaxy and the quasar \citep{Refsdal64}.
A critical aspect of this technique is a model that describes the mass distribution in the lensing galaxy and along the line of sight between the background object and the observer.  
This model is constrained by the morphology of the lensed emission of the background object, the stellar velocity dispersion in the lensing galaxy, and by deep imaging and spectroscopy of the fields containing the lens system.  
This model is combined with the time delays \citep[e.g.,][]{BonvinEtal18_PGTD} %and the microlensing effect between images \citep[e.g.,][]{TieKochanek18,GChenEtal18a} 
to measure the characteristic distances for the lens system: the angular diameter distance to the lens ($\Dd$) and the time-delay distance, 
which is a ratio of the angular diameter distances in the system: 
\begin{equation}
\label{eq:theory7}
\Ddt\equiv\left(1+
z_{\rm d}\right)\frac{\Dd\Ds}{\Dds}\propto H_{0}^{-1},
\end{equation}
where $z_{\rm d}$ is the redshift of the lens, $\Ds$ is the distance to the background source, and $\Dds$ is the distance between the lens and the source. 
In turn, these distances are used to determine cosmological parameters, primarily $H_0$ \citep[e.g.,][]{SuyuEtal14,BonvinEtal16,BirrerEtal19,GChenEtal19,RusuEtal19_H0LiCOW, WongEtal19,JeeEtal19,TaubenbergerEtal19,ShajibEtal20_0408}.

A recent analysis with this technique, using a blind analysis on data from six gravitational lens systems\footnote{Except the first lens, B1608+656, which was not done blindly, the subsequent five lenses in H0LiCOW are analyzed blindly with respect to the cosmological quantities of interest.}, inferred $H_0 = 73.3\substack{+1.7\\-1.8}~\kmsmpc$, a value that was $3.8\sigma$ offset from the Planck results \citep{WongEtal19,MillonEtal20}.  This analysis used two commonly used descriptions of the mass distribution of the lensing galaxy.  
The first description consists of a NFW halo \citep{NavarroEtal96} plus a constant mass-to-light ratio stellar distribution (the ``composite model'').  
The second description models the three dimensional total mass density distribution, i.e., luminous plus dark matter, of the galaxy as a power law \citep[][]{Barkana98}, i.e., $\rho(r) \propto r^{-\gamma}$ (the power-law model). 
These models yield $H_0$ measurements that are consistent within the errors for individual lens systems; the final uncertainties on $H_0$ incorporate a marginalization over the choice of mass model \citep{MillonEtal20}.

Although the power-law and composite models are well-motivated by both observations \citep[e.g.,][]{KoopmansEtal06,KoopmansEtal09,SuyuEtal09,AugerEtal10,BarnabeEtal11,SonnenfeldEtal13,HumphreyBuote10,Cappellari16} and simulations \citep{NavarroEtal96}, there is a well-known degeneracy in gravitational lensing known as the mass-sheet transformation (MST), that leaves imaging observables invariant but will bias the determination of $H_0$ \citep{FalcoEtal85,GorensteinEtal88}. The line-of-sight mass distribution contributes to first order mass-sheet-like effect \citep{FassnachtEtal02,SuyuEtal13,GreeneEtal13,CollettEtal13}; we will refer to this as an external MST. 
However, %it is also possible to have an effective internal MST through, 
for the mass distribution of the lensing galaxy, there are different models that can give the same lensing observables but would give different time-delays. 
The most degenerate case is the one with spherical symmetry in which the density profiles differ by a component that is uniform in within the radial ranges probed by lensing. This component which could be described by a large-core mass distribution \citep[][see detail in Section 2]{BlumEtal20}, change the distribution of the mass density profile of the lensing galaxy. It fits with recent work that question whether that elliptical galaxies do not necessarily follow a power-law or composite model to the desired precision \citep{SchneiderSluse13, XuEtal16}.

\citet{BirrerEtal20} (hereafter Paper IV) showed that allowing for an internal MST on the power-law model increased the uncertainty of the $H_{0}$ measurement of a seven-lens sample from the $2.4\%$ precision of \citet{MillonEtal20} to $8\%$, in a $\Lambda$CDM cosmology. 
Interestingly,
the central value of $H_{0}$ remained almost unchanged in this analysis ($74.5\substack{+5.6\\-6.1}~\kmsmpc$). 
To improve the precision of the $H_0$ inference, Paper IV added data from the SLACS sample \citep{BoltonEtal04SDSS,BoltonEtal06SDSS}. %under the assumption that the deflectors are drawn from the same population. (which may not be true, because, e.g. are at a different redshift, and the selection is different). 
In this lens sample, the background objects are galaxies, not quasars, so they cannot be used for time-delay cosmography. 
However, the combination of high-resolution imaging and kinematic measurements allows the SLACS sample to improve the constraints on the mass profiles of massive elliptical galaxies.  
With the inclusion of the SLACS information \citep{ShajibEtal20} and the assumption that the sample of time-delay and SLACS lenses are drawn from the same population, the inference on $H_{0}$ shifted to $67.4\substack{+4.1\\-3.2}~\kmsmpc$, in agreement with the Planck value and results from distance ladders \citep{RiessEtal19,FreedmanEtal19}.
Comparison of the galaxy population distributions show that several observed properties, such as central stellar velocity dispersion, are similar. In addition, elliptical galaxies are a very homogenous population, as evidenced by the tightness of correlations like the 
fundamental plane \citep[][and references therein]{AugerEtal10}.
% 
%On measures such as stellar velocity dispersion, the two samples appear consistent. 
However, some major differences between the samples are that (1) the SLACS lensing galaxies are at lower redshifts than those in the time-delay sample, and (2) the SLACS lensing galaxies have smaller ratio of effective radius to Einstein radius than the time-delay sample (see Fig. 16 in Paper IV).
Possible potential biases and limitations of using the SLACS sample are discussed by Paper IV and \citet{ShajibEtal20}.

In this work, we take a more general approach to constrain the internal MST by combining the time-delay lens system with relative distance indicators without assuming a specific parametrization of the cosmological model. We show that one can hence constrain the internal MST in a cosmological independent way and obtain more broadly applicable distance posteriors.
%As the distance measurements in Paper IV relies on the $\Lambda$CDM model, those distance measurements are cosmological-model-dependent quantities.
%Because we do not yet have a sample comparable to SLACS at the redshifts of the time-delay sample, it is important to apply different methods to constrain the internal MST and understand what provides the constraining power on the internal MST in these analyses.
%In this paper we 
% use real imaging data (see \fref{fig:aohstimages}) to 
%investigate what assumptions have been made in order to break the internal MST and provide a cosmological-independent way to constrain the internal MST.
In \sref{sec:jjMST}, we introduce the basics of the mass-sheet transformation. In \sref{sec:Ddt_MST} and \sref{sec:dist_MST}, we discuss the distance measurements under the effects of  the internal and external MST. In \sref{sec:error_p}, we discuss error propagation under MST.
In \sref{sec:Constrain_internal_MST}, we provide a cosmological-model-independent way to constrain the internal MST. %\sref{sec:joint_inference}, we lay out the Bayesian inference of cosmological parameters with multiple lens systems. 
We summarize our work in \sref{sec:conclusion}.

\section{The Mass-sheet transformation}
\label{sec:jjMST}
The mass-sheet transformation is a degeneracy affecting gravitational lens systems. One can transform any projected mass distribution, $\kappa(\theta)$, into infinite sets of $\kappa_{\lambda}(\theta)$ via 
\begin{equation}
\label{eq:MST}
    \kappa_{\lambda}(\theta)=\lambda\kappa(\theta)+1-\lambda,
\end{equation}
without degrading the fit to the lensed emission \citep{FalcoEtal85}, although MST does change the source size accordingly. 
Here $\kappa({\theta})$ is a scaled two-dimensional projected mass density distribution, $\kappa(\theta) = \Sigma(\theta) / \Sigma_{\rm crit}$, where $\Sigma(\theta)$ is the mass surface density and $\Sigma_{\rm crit}$ is the lensing critical density,
\begin{equation}
\label{eq:crit}
    \Sigma_{\textrm{cr}}=\frac{c^{2}}{4\pi G}\frac{\Ds}{\Dd\Dds}.
\end{equation} 
The physical picture of MST comes from both the environment (a.k.a., an external MST, $\kappaext$) and the mass models of the lensing galaxy (a.k.a., an internal MST, $\lambdaint$).
We separate these two components of the MST because we use different observables to assess their effects.
For example, the estimation of the external MST uses weighted number counts of galaxies and/or weak gravitational lensing, based on spectroscopy and deep imaging of the field containing the lens.  
This approach has been extensively used in TDC  \citep[e.g.,][]{FassnachtEtal06,SuyuEtal10,CollettEtal13,GreeneEtal13,RusuEtal17,TihhonovaEtal18,Buckley-GeerEtal20}. Information about the internal MST is derived from high-resolution imaging and the stellar velocity dispersion of the lensing galaxy.

The theoretical version of the internal MST, i.e., a mass sheet with infinite extent, is clearly non-physical.  
Therefore, in assessing the internal MST we need to find a physical model that approximates the behavior of a mass sheet at small projected distances from the center of the lensing galaxy, but that vanishes at large radii \citep[see Fig. 7 in ][]{SchneiderSluse13}.
% However, the internal MST which exactly follow \eref{eq:MST} does not exist since
%$\lambdaint$
%is only an approximation of the internal MST
%profile degeneracy across different 3D mass profile inside the arc region
% mathematically a smooth 3D mass profile which corresponds
% to 2D projected mass density with a asymptotically non-vanished $\lambdaint$ does not exist, 
% Thus, in the real physical system, the value of $\lambdaint$ needs to be vanished at large radii \citep[see Fig. 7 in ][]{SchneiderSluse13}.
%but a 3D mass profile which approximate internal MST inside the Einstien radius where extended imaging is sensitive to does exist. 
%For the internal MST,
%H0LiCOW team used different mass profiles and velocity dispersion to exam the possible internal MST and found that 
%although a pure internal MST which follow \ref{eq:MST} does not exist, 
One example of this was proposed by \citet{BlumEtal20}, who introduced a mass profile that can satisfy these requirements. For this profile,
% which can well capture the concept of internal MST and thus 
the physical internal MST, which redistributes any specific mass profile, $\kappa(\theta)$, should be written as 
\begin{equation}
\label{eq:internal_MST}
    \kappa_{\textrm{mst,int}}(\theta)=\lambdaint\kappa(\theta)+(1-\lambdaint)\kappa_{\rm c}(\theta),
\end{equation}
where 
\begin{equation}
\label{eq:core_MST}
    \kappa_{\rm c}(\theta) =\frac{\theta_{\rm s}}{\sqrt{\theta_{\rm s}^2 + \theta^2}},
\end{equation}
and $\theta_{\rm s}$ is the scale radius. When we set $\theta_{\rm s}$\ to a large value, e.g. $10\arcsec$, $\kappa_{\rm c}(\theta)$ approximates the theoretical internal MST very well over the region of interest (Paper IV).

\begin{figure}
\includegraphics[width=\linewidth]{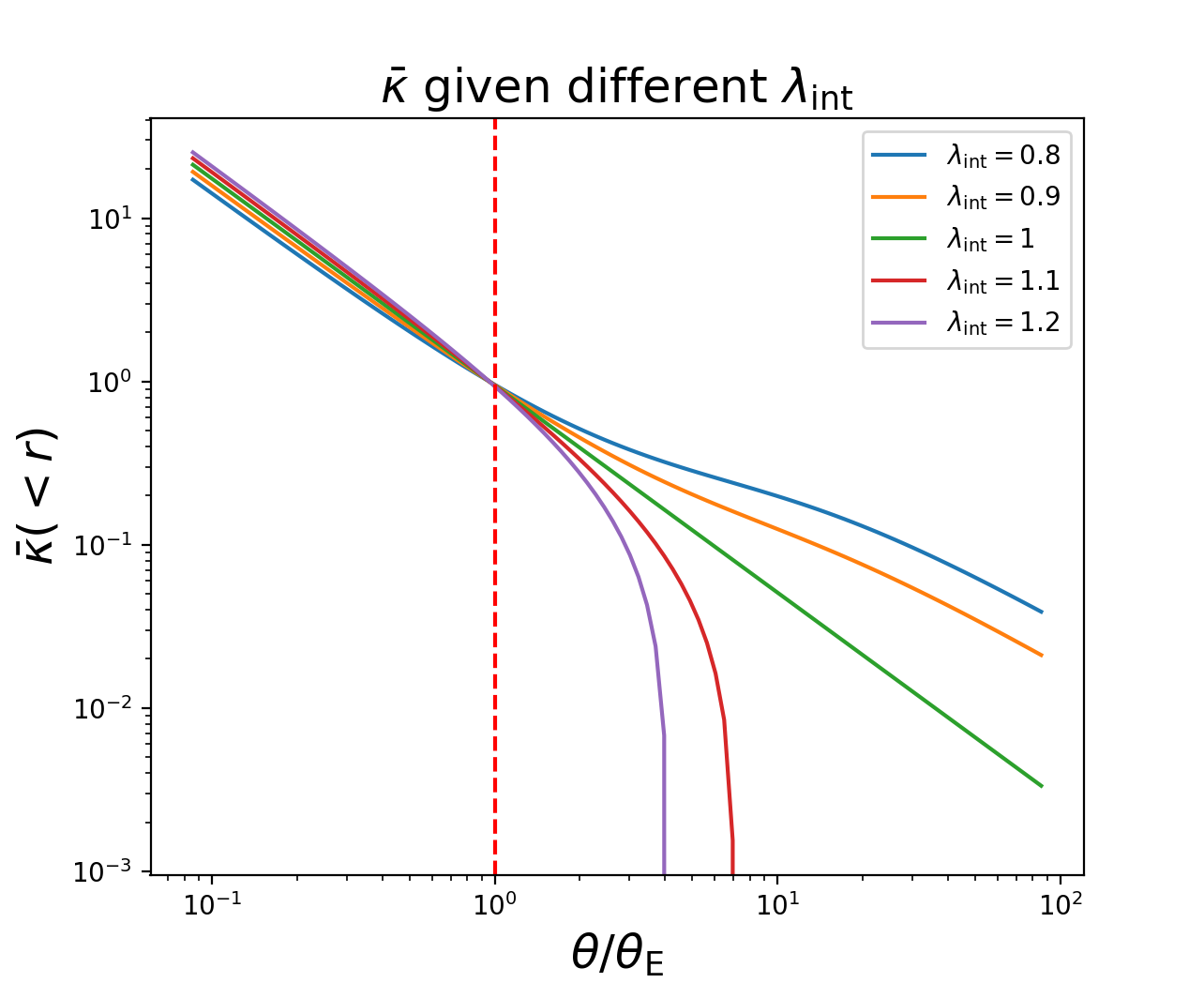} 
\caption{Illustration of the transformed power-law profile in the mean dimensionless enclosed projected mass distribution under the internal MST with $\theta_{\rm s}=10''$. All the transformed mass profiles share the same Einstein radius (red dashed line). 
All the mass distributions in this figure produce essentially the same model images but different unlensed size of the source, which is not directly observable.}
\label{fig:MST_kappa}
\end{figure}

%Similarly,
%A companion paper 
%Yildirim et al. (submitted) show that an additional mass distribution with a large core radius (e.g., $r_{\rm s}=1000\arcsec$) approximates the theoretical internal MST very well over the region of interest.
We illustrate the effects of adding such a mass-sheet profile to the lensing galaxy mass distribution in \fref{fig:MST_kappa}, by plotting the mean dimensionless enclosed projected mass distribution,
\begin{equation}
    \Bar{\kappa}(\theta)\equiv\frac{2}{\theta^{2}}\int^{\theta}_{0}\theta^{\prime}\kappa_{\textrm{mst,int}}(\theta^{\prime})d\theta^{\prime}.
\end{equation} 
The Einstein radius of the lens system, $\theta_{\textrm{E}}$, is defined as the angular radius for which $\Bar{\kappa}(\theta_{\textrm{E}})=1$.
%In such case, $\Ddt$ fully degenerate with $\lambda_{\textrm{int}}$. 

Thus, the general MST which accounts for both $\kappaext$ and $\lambdaint$ can be written as 
\begin{equation}
\label{eq:true_kappa}
    \kappa_{\lambda}(\theta)=(1-\kappaext)\kappa_{\textrm{mst,int}}(\theta)+\kappaext = \kappa_{\textrm{true}}(\theta),
\end{equation}
where $\kappa_{\textrm{true}}$ represents the ${\it true}$ projected mass profile. 
%When $\kappaext$ and $\lambdaint$ are small,
%\begin{equation}
%    1-\lambda=\kappaext+\lambdaint.
%\end{equation}
% As the mathematical form of external MST and internal MST has been well established, dissecting 
In this paper we set the stage for future investigations by dissecting where the constraining power on the distance measurements in TDC comes from, and exploring what assumptions have to be made and data have to be used in order to break the internal MST.
% are important for future studies. 

\section{The measurement of $\Ddt$ under the MST}
\label{sec:Ddt_MST}
Once the time delays between multiple images are observed, one can measure the time-delay distance via 
\begin{equation}
\label{eq:TD_Ddt}
\Delta t%=\frac{\Ddt}{c}\left[\frac{1}{2}(\theta-\beta)^{2}-\psi(\theta)\right]%,
=\frac{\Ddt}{c} \Delta\phi(\theta, \beta),
\end{equation}
where $c$ is the speed of light and $\theta$, $\beta$, and
$\phi(\theta)$ are the image coordinates, the source
coordinates, and the Fermat potential respectively. %The time-delay distance, $\Ddt$,
%contains all of the cosmological information and is defined as
%\begin{equation}
%\label{eq:theory7}
%D_{\Delta t}\equiv(1+\textit{z}_{\rm d})\frac{\Dd %D_{\rm s}}{\Dds},
%\end{equation}
%where $\Dd$, $D_{\rm s}$ and $\Dds$ are the angular diameter distances to the lens, to the source, and between the lens and the source, respectively. 
The form of \eref{eq:TD_Ddt} allows the inference of the cosmological information contained in $\Ddt$ without any need for cosmological priors on the lens modeling.
% The separability in \eref{eq:TD_Ddt} allows us to infer cosmographic information, i.e., the value of $\Ddt$, without the need of cosmological priors in the lens modeling.

However, in the presence of a MST, given the same time delays and imaging data, the transformed projected mass profile produces a different time-delay distance via
\begin{equation}
\label{eq:MST_2}
    D_{\Delta t,\lambda}=\frac{\Ddt}{\lambda}.
\end{equation}
Thus, additional information is required to constrain both the internal and external MST, and thus to obtain unbiased $\Ddt$ measurements. 

\section{The measurement of $\Dd$ under the MST} 
\label{sec:dist_MST}
Once the velocity dispersion of the lensing galaxy is measured, one can use high-resolution imaging of the lens system to measure the ratio $\DsDds$ via
\begin{equation}
\label{eq:vd_nolambda_nokappa}
    (\sigma_v^{\textrm{p}})^{2}=\left(\frac{\Ds}{\Dds}\right)c^{2}J(\eta_{\textrm{lens}},\eta_{\textrm{light}},\beta_{\textrm{ani}}),
\end{equation}
where $\sigma_v^{\textrm{p}}$ is the line-of-sight luminosity-weighted velocity dispersion that is predicted by the mass distribution in the lensing galaxy. 
Here, $J$ contains the angular-dependent information
including the parameters describing the 3D deprojected mass distribution, $\eta_{\textrm{lens}}$, the surface-brightness distribution in the lensing galaxy, $\eta_{\textrm{light}}$, and the stellar orbital anisotropy distribution, $\beta_{\textrm{ani}}$. 
In a similar way to the time-delay distance, the separability in \eref{eq:vd_nolambda_nokappa} allows us to infer the cosmological distance ratio $\DsDds$ without the need of cosmological priors on $J$. 
Since $\Ddt \propto \Dd (\DsDds)$, we can use the combination of the $\Ddt$ measurement from the time delays and $\DsDds$ from velocity dispersion to obtain $\Dd$.

We discuss the effect of $\kappaext$ and $\lambdaint$ on the $\Dd$ measurement in the following two sections.

\subsection{External MST only}
\citet{JeeEtal15} found that $\Dd$ is an invariant quantity under an external MST.
%The measurement of $\Dd$ can be obtained by combining time delays, imaging data, and the velocity dispersion and its value is invariant when $\lambda$ only accounts for $\kappaext$ (Jee et al. 2015, Jee et al. 2019). 
This is because $\kappaext$ only contributes to the change of the normalization of 3D mass profile and does not affect its overall shape given any mass model \citep[][]{SuyuEtal13,GChenEtal19}. 
That is,
% Due to the fact that $\kappaext$ only change the normalization of 3D mass profile,
\eref{eq:vd_nolambda_nokappa} changes to 
\begin{equation}
\label{eq:vd_nolambda}
    (\sigma_v^{\textrm{p}})^{2}=(1-\kappaext)\left(\frac{\Ds}{\Dds}\right)_{\kappaext} c^{2}J(\eta_{\textrm{lens}},\eta_{\textrm{light}},\beta_{\textrm{ani}}),
\end{equation}
where the minus sign means that one needs to remove the mass contributed from the environment (i.e., the mass along the line of sight) inside the Einstein radius to obtain the mass of the lensing galaxy. 

\begin{figure}
\centering
    \includegraphics[width=0.7\linewidth]{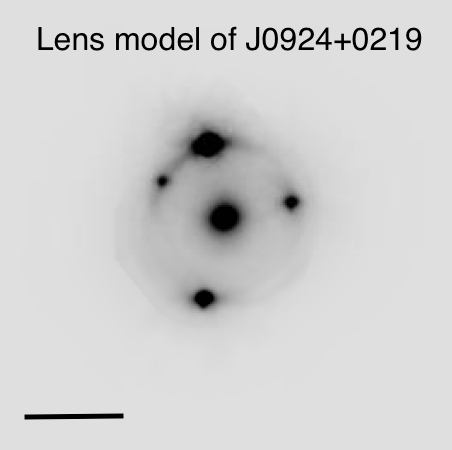}
    \caption{The multiple lensed images and the extended arc around the lensing galaxy are from the background AGN and its reconstructed host galaxy. The foreground main lens is located in the center of the lens system. 
    The solid horizontal line represents 1\arcsec scale.
    The detailed lens modeling will be presented in Chen et al. in prep.
    Note that the study of this paper is not limited to any specific configuration of the lens systems.}
\label{fig:aohstimages}
\end{figure}

\begin{figure*}
\includegraphics[width=\linewidth]{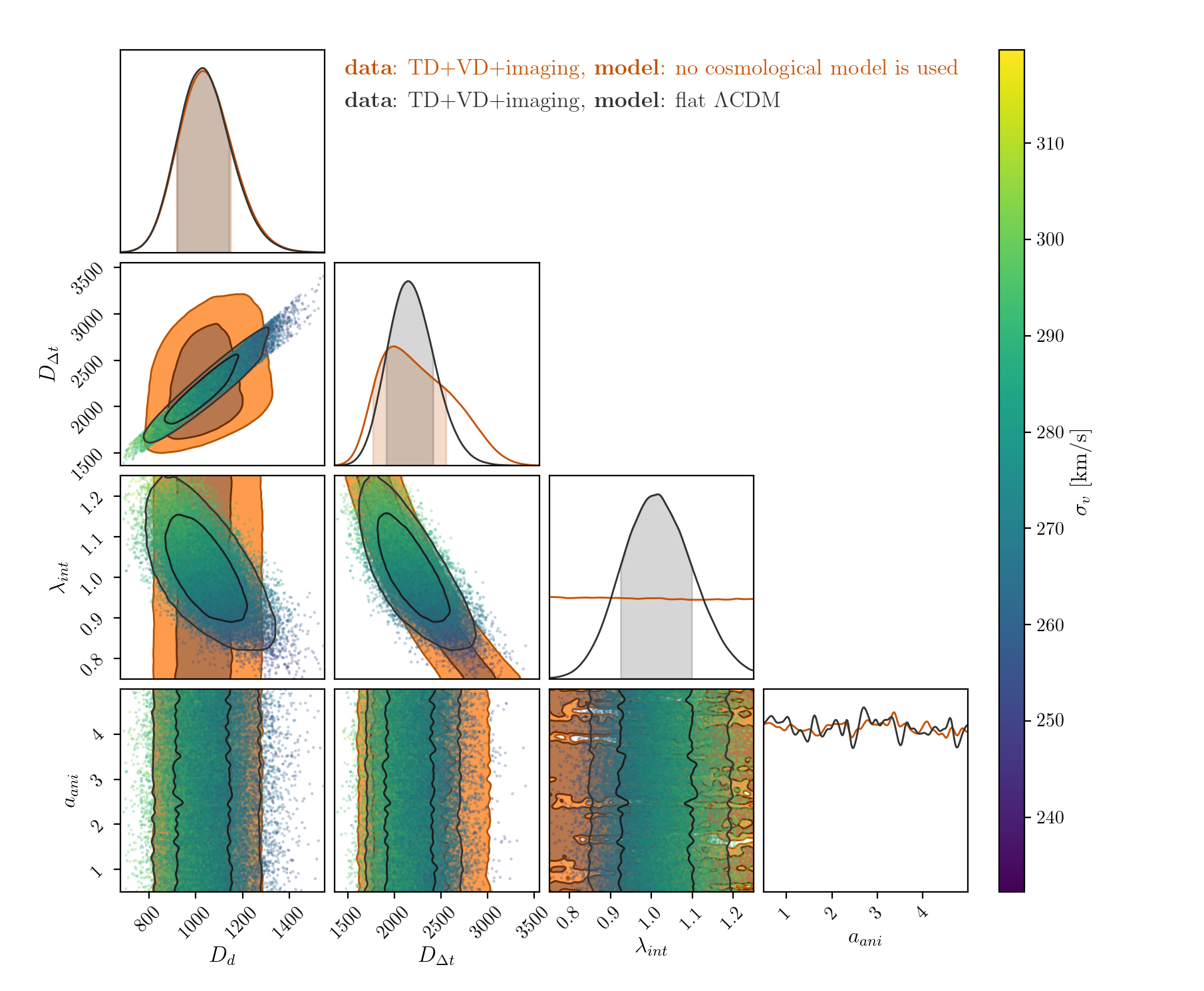} 
\caption{The comparison of the $\Dd$, $\Ddt$, $\lambdaint$, and $a_{\textrm{ani}}$ measurements with and without the assumption of $\Lambda$CDM model from single time-delay mock lens.
`TD' represents time delay information and `VD' represents velocity dispersion information. When $\Lambda$CDM model is not assumed, the internal MST ($\lambdaint$) is not constrained.
When $\Lambda$CDM model is assumed, the degeneracy can be broken and hence $\Ddt$ is constrained. 
The anisotropy parameter is not constrained in either case. 
%Note that when $\Lambda$CDM is used, we assume $\Omega_{\textrm{m}}=[0.05,1]$, $\Omega_{\Lambda}=1-\Omega_{\textrm{m}}$, and $H_{0}$ uniform in $[0,150]~\kmsmpc$.
Color-coded velocity dispersion shows that $\Ddt$ is positively correlated with $\Dd$ but anti-correlated with $\sigma_v$. The contours represent the 68.3\% (shaded region) and 95.4\% quantiles.}
\label{fig:cosmology_compare}
\end{figure*}

In order for the predicted velocity dispersion to match the observed value, $\DsDds$ must transform under an external MST via
\begin{equation}
\label{eq:MST_DsDds}
    \left(\frac{\Ds}{\Dds}\right)_{\kappaext}=(1-\kappaext)^{-1}\left(\frac{\Ds}{\Dds}\right)=\left(\frac{\Ds}{\Dds}\right)_{\textrm{true}}.
\end{equation}
Then the time-delay distance changes via
%while time-delay constrain the the product of $\Ds/\Dds$ and $\Dd$ and it transform under external MST via
\begin{equation}
\label{eq:MST_Ddt}
    D_{\Delta t,\kappaext}=(1-\kappaext)^{-1}\Ddt.
\end{equation}
Thus, by combining \eref{eq:theory7}, \eref{eq:MST_DsDds}, and \eref{eq:MST_Ddt}, one can show that $\Dd$ is invariant under the external MST,
\begin{equation}
    (\Dd)_{\kappaext} = \Dd.
\end{equation}

%We illustrate it in \fref{fig:vary_kappa}.

%\begin{figure}
%\includegraphics[width=\linewidth]{MST_enclosedmass.png} 
%\caption{Illustration of the 3D enclosed mass of the power-law profile under an approximate MST with a cored mass component (\eref{eq:core_MST}).}
%\label{fig:MST_enclosedmass}
%\end{figure}

\begin{figure*}
\includegraphics[width=\linewidth]{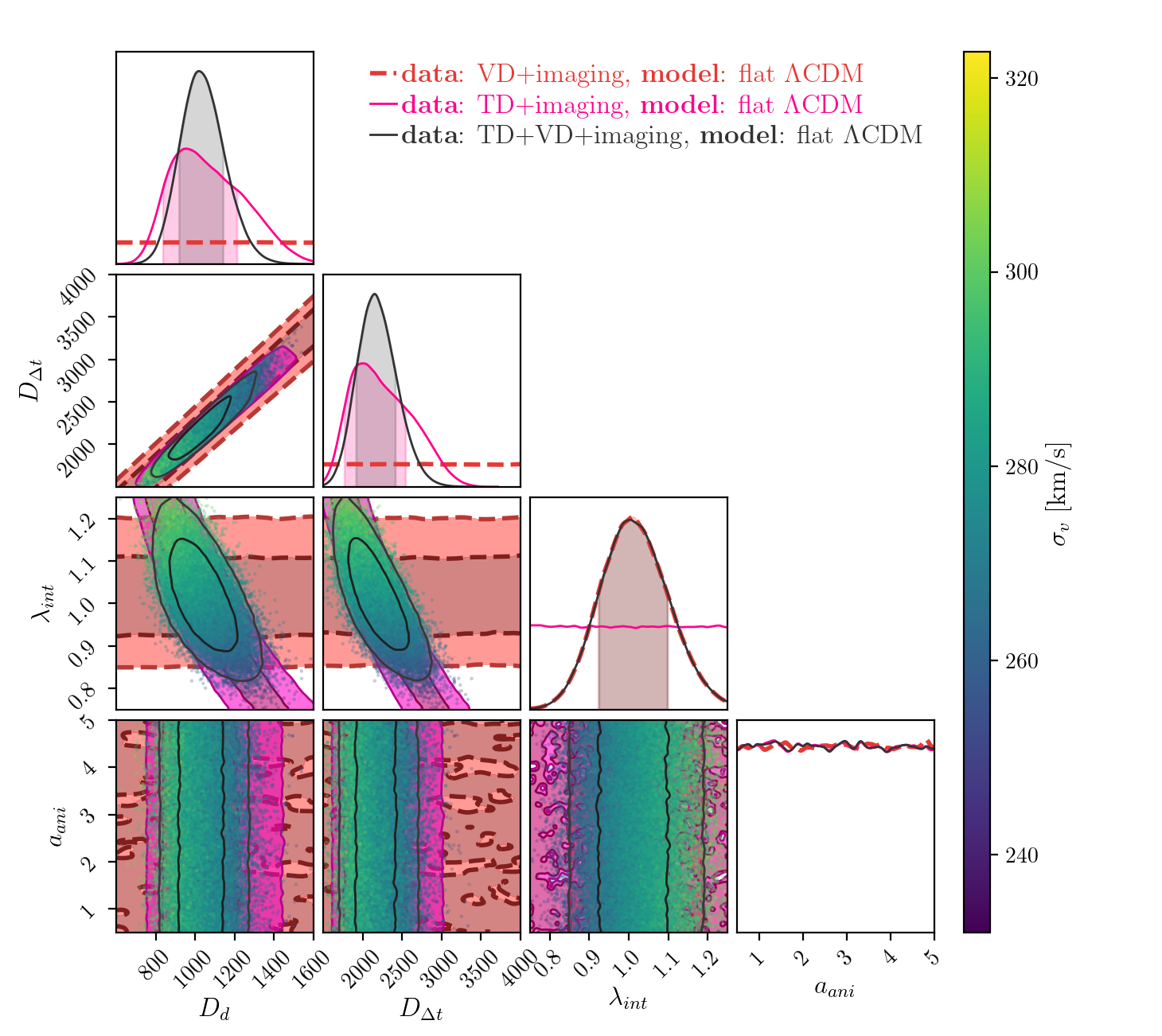}
\caption{The decomposition of the constraining power from time delays (TD), velocity dispersion (VD), and imaging data under the assumption of $\Lambda$CDM model. 
%posteriors of the measurements of $\Dd$, $\Ddt$, $\lambdaint$, and $a_{\textrm{ani}}$ 
When only `TD+imaging' is used, the values of $\Dd$ and $\Ddt$ are not constrained, and both of them are fully degenerate with $\lambdaint$. 
Since the velocity dispersion constrains the $\lambdaint$, the joint constraint (`TD+VD+imaging data') breaks the degeneracy and hence constrains $\Ddt$ and $\Dd$. The anisotropy parameters are not constrained in all cases.}
\label{fig:cosmology_LCDM}
\end{figure*}

%\begin{figure}
%    \includegraphics[width=\linewidth]{vary_kext.png}
%    \caption{Illustration of the invariant $\Dd$ measurements under different value of $\kappaext$.}
%\label{fig:vary_kappa}
%\end{figure}

\subsection{Internal MST+External MST}
\label{sec:internal_MST}
Since the velocity dispersion depends on the enclosed 3D mass of the lensing galaxy, whose shape is not conserved under an internal MST, the constraint on the $\Ds/\Dds$ ratio is not mathematically scaled by $\lambdaint^{-1}$ under an internal MST. 
We therefore must expand the \eref{eq:vd_nolambda} by including $\lambdaint$ in the $J$ term,
\begin{equation}
\label{eq:vd_withlambda}
    (\sigma_v^{\textrm{p}})^{2}=(1-\kappaext)\left(\frac{\Ds}{\Dds}\right)_{\textrm{true}}J(\eta_{\textrm{lens}},\eta_{\textrm{light}},\beta_{\textrm{ani}},\lambdaint),
\end{equation}
since $J$ contains the 3D de-projected dimensionless mass model components, whose structure is affected by the value of $\lambdaint$. Thus, $\Dd$ can be expressed as 
\begin{equation}
\label{eq:Dd_lambbdaint}
    \Dd=\frac{1}{1+z_{\rm d}}\frac{\Ddt}{\lambdaint}\frac{c^{2}}{\sigma_v^2}J(\eta_{\textrm{lens}},\eta_{\textrm{light}},\beta_{\textrm{ani}},\lambdaint).
\end{equation}
%We use mock time delays assuming $t_{\Delta AB}=10\pm1.5$ days, $t_{\Delta AB}=15\pm1.5$ days, $t_{\Delta AB}=10\pm1.5$ days, and mock velocity dispersion measurements ($=280\pm25\kms$) to constrain $\Ds/\Dds$ and $\Ddt$ under the internal MST without assuming any cosmological model. 

%Many previous investigations claimed that the internal MST can be broken by having additional information from single-aperture velocity dispersion. 
Many previous investigations \citep[e.g.][Paper IV]{SuyuEtal14} show that the internal MST can be broken with a single-aperture velocity dispersion, given a cosmological model.
%The argument behind the idea was that the extended lensed arcs constrain the mass inside the Einstein radius, while the velocity dispersion constrains the mass inside the effective radius \citep[see also][]{DuttonEtal11}. 
%These two measurements at different radii constrain the slope and hence limit the range of the internal MST. 
%We want to stress that both velocity dispersion and cosmological model assumption are necessary to break the internal MST. 
This can be explained as follows: firstly, the Einstein ring radius, as defined in terms of the mean dimensionless enclosed projected mass distribution ($\Bar{\kappa}$), is invariant under an internal MST (i.e., $\Bar{\kappa}(\theta_{\textrm{E}}) =\Bar{\kappa}_{\lambda}(\theta_{\textrm{E}})=1$; see also \fref{fig:MST_kappa}), while the physical mass inside the Einstein radius is unconstrained without assuming a cosmological model.  
%the extended arc can only constrain the mean dimensionless enclosed projected mass distribution inside Einstein radius. That is, $\Bar{\kappa}(\theta_{\textrm{E}}) =\Bar{\kappa}_{\lambda}(\theta_{\textrm{E}})=1$, where $\Bar{\kappa}(\theta)\equiv\frac{2}{\theta^{2}}\int^{\theta}_{0}\theta^{\prime}\kappa(\theta^{\prime})d\theta^{\prime}$ and hence $\theta_{\textrm{E}}$, the definition of Einstein radius, is invariant under internal MST (see also \fref{fig:MST_kappa}), while the physical mass inside the Einstein radius is still unconstrained without assuming a cosmological model. %as $\Dd$, which is needed to assign the mass inside $\theta_{e}$, is a free valuable. 
Secondly, from \eref{eq:vd_withlambda} we show hereafter that if a cosmological model is not assumed, then the values of $\lambdaint$ and $\DsDds$, which affect the shape and normalization respectively, are degenerate.
Hence, even with a measured velocity dispersion, the mass inside the effective radius is also not constrained. 
Therefore, a single-aperture velocity dispersion is insufficient to break the degeneracy and constrain the internal MST if one does not assume a cosmological model. Spatially resolved kinematics of the lensing galaxy would be required (Yildirim et al. in prep.).

In order to illustrate these dependencies, we use the power-law mass model, which was obtained by fitting to the real imaging data of a four-image gravitational lens system (J0924+0219) shown in \fref{fig:aohstimages} (see Chen et al.\ in prep for details), and analysed it in the context of an internal MST (i.e., added to our model a MST component as described in  \eref{eq:core_MST}).
For the anisotropy component, we assume $\beta_{\rm ani}$ varies with radius, and parameterize this behaviour in the form of an anisotropy radius, $r_{\textrm{ani}}$, in the Osipkov-Merritt formulation \citep{Osipkov79,Merritt85}, 
\begin{equation}
    \beta_{\textrm{ani}}=\frac{r^{2}}{r^{2}_{\textrm{ani}}+r^{2}},
\end{equation}
as an example.
In this formulation, $r_{\textrm{ani}}=0$ indicates pure radial orbits and $r_{\textrm{ani}}\rightarrow\infty$ is isotropic with equal radial and tangential velocity dispersions. 
In our models, we use a scaled version of the anisotropy parameter, $a_{\textrm{ani}}\equiv r_{\textrm{ani}}/r_{\textrm{eff}}$, where $r_{\textrm{eff}}=\Dd \theta_{\textrm{eff}}$, and $\theta_{\textrm{eff}}$ is the effective radius.
The redshift of the lens and source are $z_{\rm d}=0.393$ and $z_{\rm s}=1.523$, respectively. 
Note that the study of this work is not limited to any specific configuration of the lens systems. We set mock time delays ($\Delta t_{\rm AB}=10\pm1.5$ days, $\Delta t_{\rm CB}=15\pm1.5$ days, $\Delta t_{\rm DB}=10\pm1.5$ days), and a mock velocity dispersion measurement ($=279\pm15~\kms$) for the analysis. The uncertainties on the time delays and velocity dispersion are typical of time-delay lens systems. 
Note that all the figures produced in this work are based on this single lens. Since $\kappaext$ does not affect the $\Dd$ measurement and is well-understood, we set $\kappaext=0$ throughout the paper for simplicity.

We consider two situations, one with a flat $\Lambda$CDM cosmology, and another in which we only use the velocity dispersion, time delays, and imaging data without assuming any cosmological model. Note that throughout the paper, for flat $\Lambda$CDM cosmology, we assume $\Omega_{\textrm{m}}=[0.05, 1.0]$, $\Omega_{\Lambda}=1-\Omega_{\textrm{m}}$, and $H_{0}$ uniform in $[0,150]~\kmsmpc$.
The results are shown in \fref{fig:cosmology_compare}, where the $\Lambda$CDM results are shown as points that are color-coded by the velocity dispersion in order to demonstrate the link between $\sigma_v$ and the other parameters. \fref{fig:cosmology_compare} clearly demonstrates that $\lambdaint$ can be constrained only when $\Lambda$CDM is assumed,
%In \fref{fig:cosmology_compare}, we show that $\Dd$ is not mathematically invariant under internal MST by using mock time delays and velocity dispersion\footnote{We use J0924+0219 HST imaging and AO imaging from the SHARP program to constrain the power-law mass profile, but create three mock time delays $t_{\Delta BA}=10\pm1.5$ days, $t_{\Delta CA}=15\pm1.5$ days, $t_{\Delta DA}=10\pm1.5$ days, and mock velocity dispersion measurement ($=280\pm25~\kms$) for demonstrating the concept of this paper.}.
%As Paper IV found that the physical boundary condition of $\lambdaint$ is inside the range $=[0.8, 1.2]$ for the power-law model, we set only a bit larger range of $\lambdaint$ to $[0.75, 1.25]$ for better visualisation.
%We can see that $\lambdaint$ is not constrained in the ``free cosmology'' scenario. 
%while, within the physically-allowed values of $\lambdaint$, the value of $\Dd$ changes very little as $\lambdaint$ changes. 
on the contrary, when the cosmological model constraint is relaxed, $\Dd$ changes very little  within the physically-allowed values of $\lambdaint$. 
Thus the $J$ term in \eref{eq:vd_withlambda} can be very well approximated as $J(\lambdaint)=\lambdaint J$ with $<1\%$ shift on the distance measurement of $\Dd$ for $\theta_{\rm s}=10''$ and with single aperture averaged velocity dispersion. This approximation was also used by Paper IV.
%Thus, we can very well approximate \eref{eq:vd_withlambda} as 
%\begin{equation}
%    (\sigma_v^{\textrm{p}})^{2}=(1-\kappaext)\left(\frac{\Ds}{\Dds}\right)_{\textrm{true}}\lambdaint J(\theta_{\textrm{eff}},\eta_{\textrm{lens}},\eta_{\textrm{light}},\beta_{\textrm{ani}}).
%\end{equation}

% Although $\lambdaint$ is not constrained, it is important to note that given the nearly vertical degeneracy between $\lambdaint$ and $\Dd$, the value of $\Dd$ vary little inside the range of $\lambdaint =[0.8,1.2]$. That is, $\Dd$ is insensitive to the internal MST.

%\section{The main source of the cosmological information under internal MST}
%When adopting two different mass profiles to accounts for internal MST, H0LiCOW collaboration shows that $\Ddt$ provides most of the cosmological information
%\citep[e.g.,][]{BonvinEtal17,RusuEtal19_H0LiCOW,GChenEtal19,BirrerEtal19,WongEtal19}. In this section,
%we exam what's the main source of the cosmological information in the presence of internal MST by imposing flat prior on $\lambdaint =[0.8,1.2]$ that is permissible based on constraints on galaxy profiles \citet{BirrerEtal19}.

%We then use $\Ddt$ and $\Dd$ to anchor SN dataset to obtain $H_{0}$ measurement. The constraining power on the value of $H_{0}$ can show 

\section{Error propagation in MST}
\label{sec:error_p}
\subsection{Error propagation without assuming a cosmological model}
In the previous section, we showed that $\Ddt$ is directly affected by both external and internal MST, while $\Dd$ is not affected by $\kappaext$ and is nearly invariant. 
Thus, based on \eref{eq:MST_2} the error on $\Ddt$ ($\delta \Ddt$) given $\lambda$ and $\lambdaint$ scales as
\begin{equation}
\label{eq:error_Ddt_lambda}
    \frac{\delta \Ddt}{\Ddt}\sim -\frac{\delta \lambda}{\lambda}=-\frac{\delta \lambdaint}{\lambdaint},
\end{equation}
while based on \eref{eq:Dd_lambbdaint}, the error of $\Dd$ scales as 
\begin{equation}
\label{eq:error_Dd_vd}
    \frac{\delta \Dd}{\Dd}\sim -2\frac{\delta \sigma_v}{\sigma_v}, 
\end{equation}
where $\sigma_v$ is the measured line-of-sight velocity dispersion.
Thus, the uncertainties on $\Dd$ are dominated by the velocity dispersion measurement errors, while the $\Ddt$ uncertainties are dominated by both internal and external MST. Therefore, $H_0$ inferred solely from $\Dd$ is robust against the MST \citep{JeeEtal19}\footnote{Note that \citet{JeeEtal19} shows that different anisotropy model may slightly shift the inferred $\Dd$ value.}. %Note that internal MST is unconstrained if free cosmology is assumed.

\subsection{Error propagation under $\Lambda$CDM model}

%Although we have demonstrated that internal MST cannot be constrained by velocity dispersion and hence $\Ddt$ is not constrained either, 
However, if one assumes a $\Lambda$CDM model, the error on $\lambdaint$ is 
\begin{equation}
\label{eq:error_kint_vd}
    \frac{\delta \lambdaint}{\lambdaint}\sim2\frac{\delta \sigma_v}{\sigma_v}.
\end{equation}
%If we neglect the uncertainties on $\kappaext$ in \eref{eq:error_Ddt_lambda}, which is reasonable in the presence of an internal MST, 
By combining Equations \ref{eq:error_Ddt_lambda}--\ref{eq:error_kint_vd},
%, \eref{eq:error_Dd_vd}, and \eref{eq:error_kint_vd}, 
the correlations between the errors on $\Ddt$, $\Dd$, and $\sigma_v$ are
\begin{equation}
\label{eq:Ddt_Dd_sigma}
    \frac{\delta \Ddt}{\Ddt}\sim \frac{\delta \Dd}{\Dd}\sim-2\frac{\delta\sigma_v}{\sigma_v}.
\end{equation}
%under $\Lambda$CDM model. 
In \fref{fig:cosmology_compare}, we indeed see that under the assumption of $\Lambda$CDM model, $\Ddt$ is positively correlated with $\Dd$ but anti-correlated with $\sigma_v$. 

Most importantly, while \eref{eq:error_kint_vd} and \eref{eq:Ddt_Dd_sigma} tell us that both $\Ddt$ and $\Dd$ are anti-correlated with $\lambdaint$ under $\Lambda$CDM model, $\Dd$ is nearly uncorrelated with $\lambdaint$ without assuming a cosmological model inside the physically-allowed range of $\lambdaint$.

\section{Constraining the internal MST}
\label{sec:Constrain_internal_MST}
% As the internal MST is currently the dominant source of uncertainties in TDC, it is critically important to understand how to constrain the internal MST and what assumptions have to be made. 
%In this section, we summarize the different methods to break the internal MST either by cosmologically-dependent way or cosmologically-independent way. 
In the previous sections of this paper, we have shown that in the presence of an internal MST, parameterized by $\lambdaint$, the time-delay distance measurement is poorly constrained unless a specific cosmological model is picked.  This situation is clearly demonstrated in the $\lambdaint$ vs.\ $\Ddt$ panel in \fref{fig:cosmology_compare}, where $\lambdaint$ is essentially unconstrained without the assumption of a cosmological model.  
In turn, the large uncertainties in $\lambdaint$ translate into imprecise inferences on $\Ddt$.
Therefore, in this section we describe two approaches to improving the constraints on $\lambdaint$.
The first approach constrains $\lambdaint$ in a fashion that depends on the cosmological model (Paper IV), while the second approach works even if we are agnostic about the cosmological model.
In both cases, we assume that observations have provided measurements of time delays and luminosity-weighted stellar velocity dispersions with errors that are typical of those found in previous works in this field.

\begin{figure}
\includegraphics[width=\linewidth]{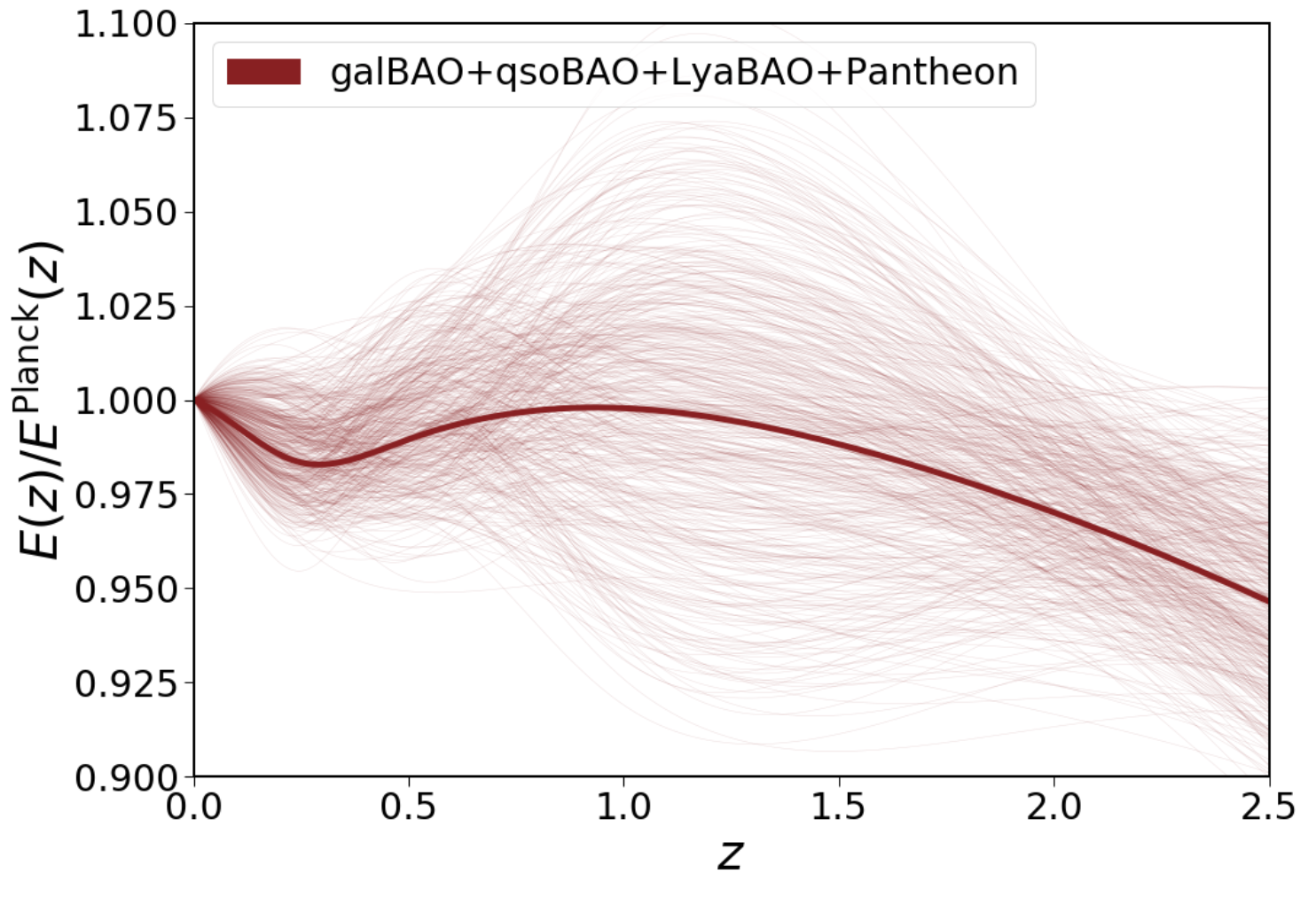}
\caption{Results of the reconstruction of $E(z)$ ($=H(z)/H_{0}$) using the SN1a and BAO data, normalised by the values for our fiducial cosmology given by the best-fitting parameters from the Planck analysis for a $\Lambda$CDM model. Each thin line comes from a random draw among the points in parameter space within 68\% confidence level.
The thick line in the middle represents the best fit. The shape of expansion history is described by piece-wise natural cubic splines. The splines can be used to constrain $\DsDds$ and hence break the internal mass-sheet transformation.}
\label{fig:cubic_JLA}
\end{figure}

\begin{figure*}
\includegraphics[width=\linewidth]{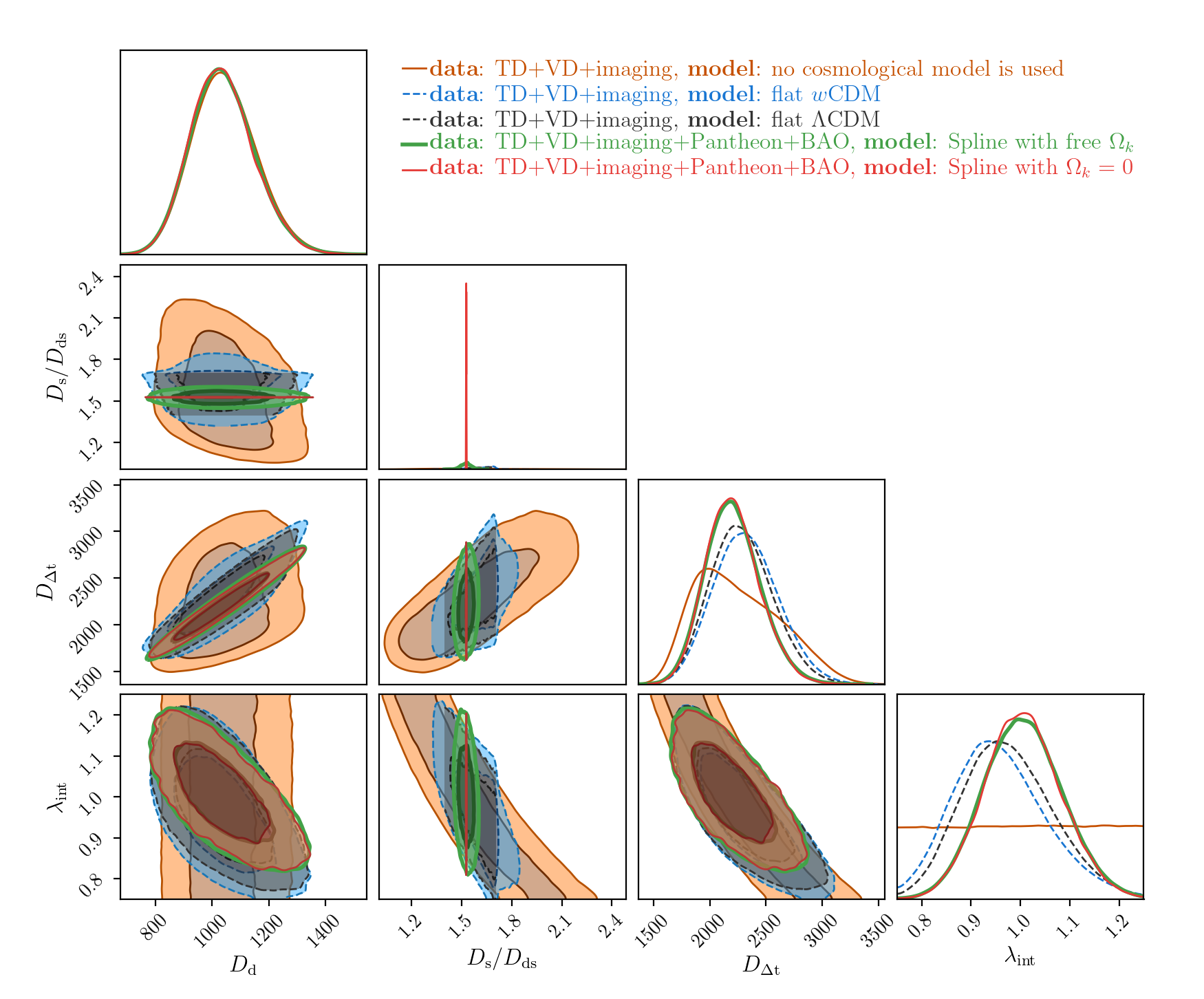}
\caption{Comparison of the inferred $\Dd$, $\DsDds$, $\Ddt$, and $\lambdaint$ in different cases: (1; orange) The distance measurements directly from single time-delay mock lens without assuming any cosmological model. 
(2; blue) The distance measurements under the assumption of flat $w$CDM model with $\Omega_{\textrm{m}}=[0.05,1.0]$, $\Omega_{\rm de}=1-\Omega_{\textrm{m}}$, $w=[-2.5,0.5]$, and $H_{0}$ uniform in $[0,150]~\kmsmpc$.
(3; black) The distance measurements under the assumption of $\Lambda$CDM model with $\Omega_{\textrm{m}}=[0.05,1.0]$, $\Omega_{\Lambda}=1-\Omega_{\textrm{m}}$, and $H_{0}$ uniform in $[0,150]~\kmsmpc.$
(4; green) The distance measurements from single time-delay mock lens, Pantheon dataset, and BAO dataset by using splines with free $\Omega_{k}$. 
(5; red) The distance measurements from single time-delay mock lens, Pantheon dataset, and BAO dataset by using splines with $\Omega_{k}=0$. 
%When combining SN1a and BAO dataset, the cosmological-model-independent $\Dd$ and $\Ddt$ measurements are comparable to those obtained by assuming a $\Lambda$CDM cosmology. 
For the cases without assuming particular cosmological model, the constraining power on $\lambdaint$ is comparable to the cases with an assumption of having a underlying cosmological model.
}
\label{fig:Dd_JLA_Ddtprior}
\end{figure*}

\begin{figure*}
\includegraphics[width=\linewidth]{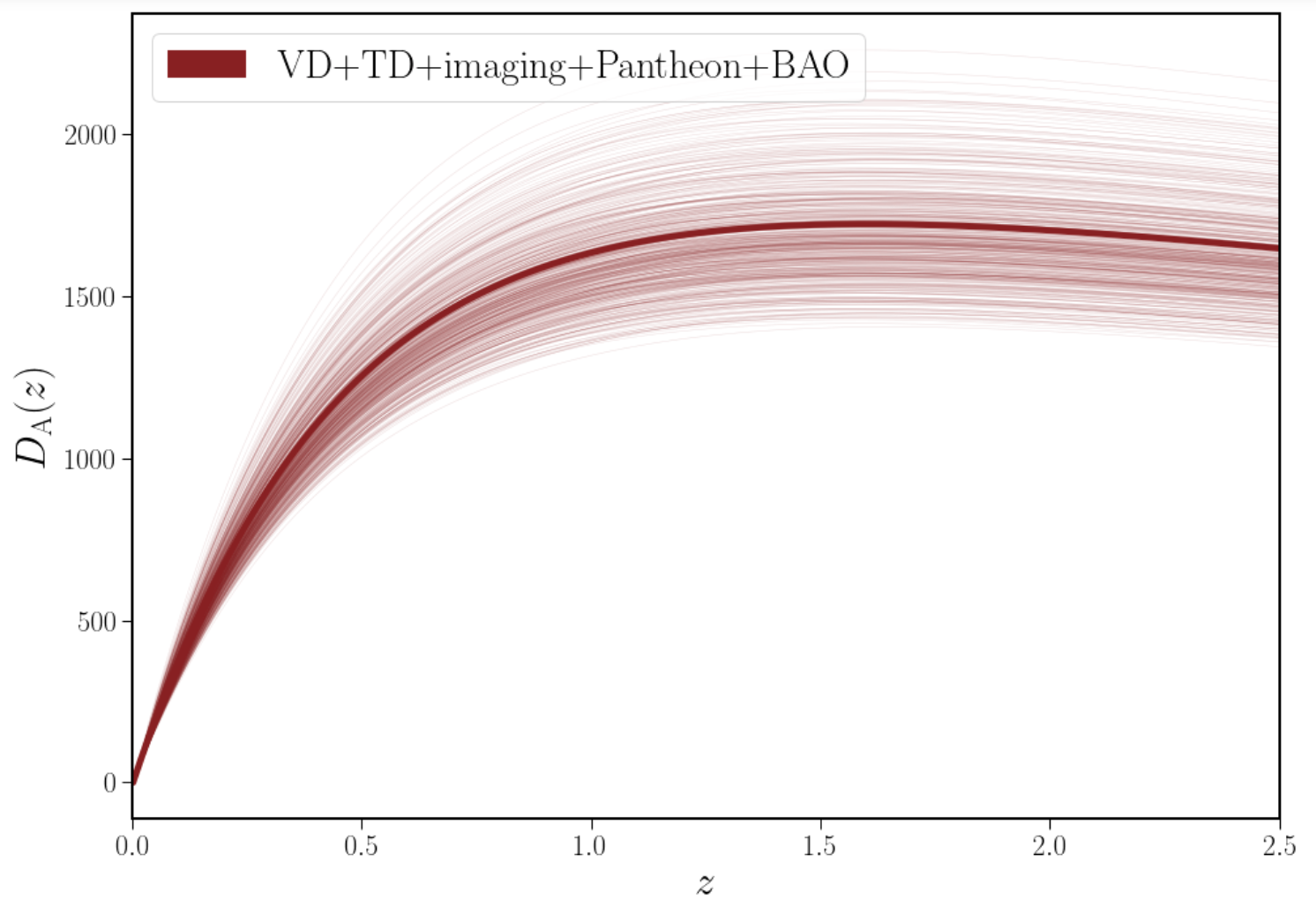}
\caption{The cosmological-model-independent distance measurements from combining single time-delay mock lens with Pantheon and BAO datasets. Each thin line comes from a random draw among the points in parameter space within 68\% confidence level. The thick line in the middle represents the best fit.}
\label{fig:hz_recon}
\end{figure*}

\begin{figure*}
    \centering
    \includegraphics[width=1.0\textwidth]{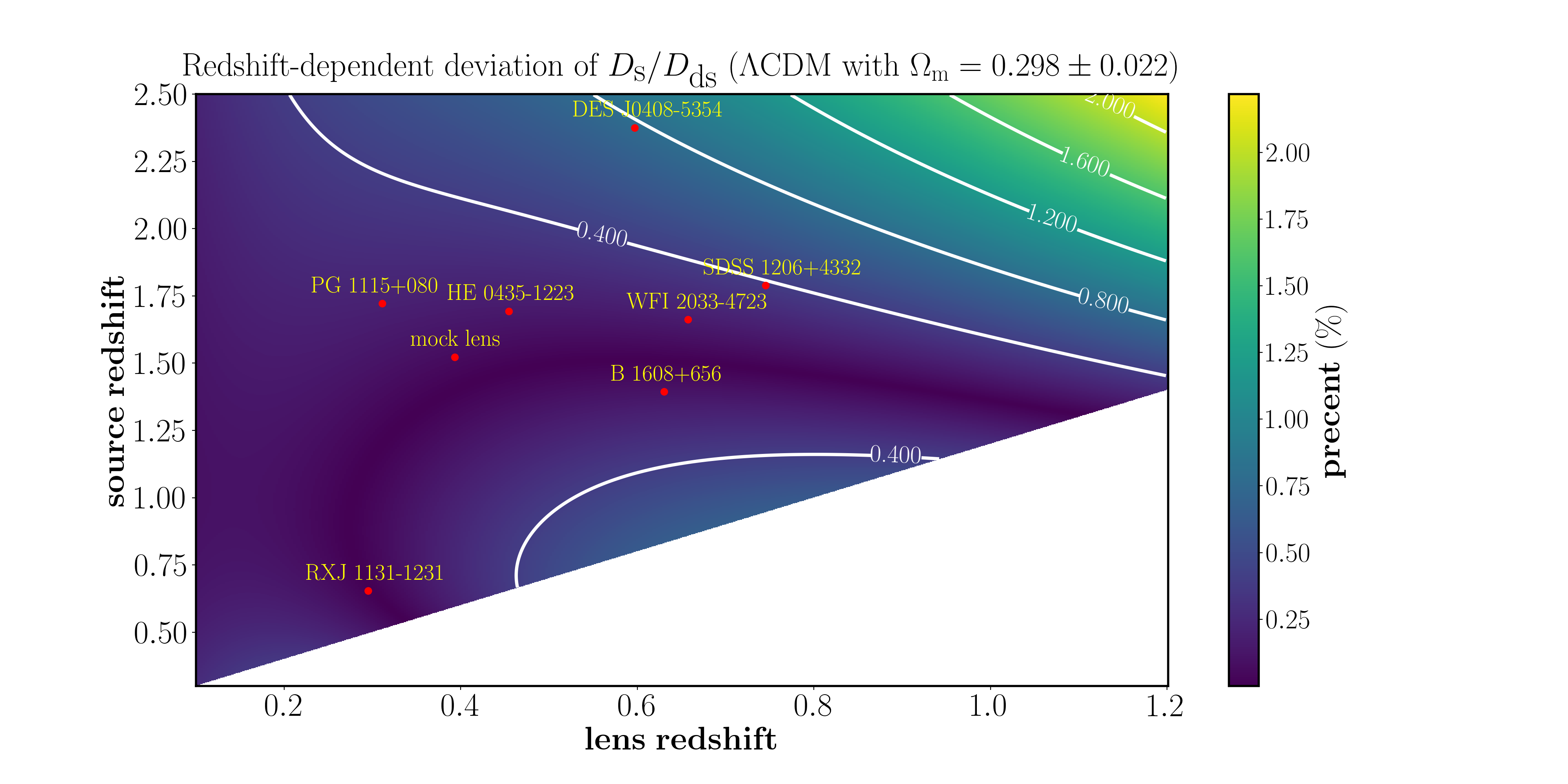}
    \caption{The percent difference between the median value of $\DsDds$, directly constrained from the SN1a and BAO, and the $\DsDds$ under the assumption of
    $\Lambda$CDM model with $\Omega_{\textrm{m}}=0.298\pm0.022$ \citep{ScolnicEtal18}, which was used in Paper IV. All 7 time-delay lenses show $<1\%$ deviation validate the use of the Pantheon data sets which constrain the low redshift expansion history and the use of $\Lambda$CDM model which extrapolates the constraint on $\DsDds$ at high redshift.}
    \label{fig:deviation_LCDM_omegaPantheon}
\end{figure*}

%\begin{figure}
%    \centering
%    \subfigure{\includegraphics[width=0.24\textwidth]{deviation_DsDds.png}}
%    \subfigure{\includegraphics[width=0.24\textwidth]{test_rasterization.png}} 
%    \caption{(a) blah (b) blah (c) blah (d) blah}
%    \label{fig:foobar}
%\end{figure}

% \subsection{Constraining $\lambdaint$ with a cosmologically-dependent way}
\subsection{Method 1: Choosing a cosmological model}
\label{sec:kappaint_cosmo_dependent}
In \sref{sec:internal_MST}, we demonstrated that $\DsDds$ and $\lambdaint$ are degenerate quantities.
However, once a cosmological model is assumed, $\DsDds$ can be determined up to a range depending on the other cosmological parameters such as $\Omega_{m}$ and $\Omega_{k}$ \citep{GrilloEtal08}, since the unknown factor of $H_0$ cancels out in the ratio. 
%Once a cosmological model is assumed, $\DsDds$ are not free variables, 
%$\DsDds$ is constrained under an assumption of a cosmological model and hence velocity dispersion can constrain $\lambdaint$. 
This means that the measurement of the velocity dispersion constrains $\lambdaint$, hence the mass inside the effective radius of the lensing galaxy.
% The mass inside the effective radius is then assigned. 
Once $\lambdaint$ is constrained, $\Ddt$ is constrained and $\Dd$ can be inferred from $\Ddt$. Hence the mass inside the Einstein radius is assigned.
Therefore, by assuming a cosmological model, the internal MST can be broken to a level that depends on the precision of the velocity dispersion measurement. 
To further illustrate the effect of assuming a cosmological model, we show the constraining power on $\Dd$, $\Ddt$, $\lambdaint$, and the anisotropy parameter ($a_{\textrm{ani}}$) when setting the cosmological model to $\Lambda$CDM in \fref{fig:cosmology_compare}. We clearly see the correlation between $\Ddt$ and inferred $\Dd$ under the assumption of $\Lambda$CDM.  %Note that Paper IV use the prior on $\Omega_{\rm m}$ based on the Pantheon samples \citep{ScolnicEtal18}

For the case where no cosmological model is used in \fref{fig:cosmology_compare}, we see that $\Ddt$ is degenerate with $\lambdaint$. In contrast, by assuming a cosmological model, one restricts the allowed range of $\lambdaint$ and thus places stronger constraints on the inferences on the cosmological distances. %This is because $\DsDds$ and $\Dd$ can be inferred from $\Ddt$ under a cosmological model and thus break the degeneracy between $\Dd$ and $\lambda_{\textrm{int}}$. 
We can decompose the black contour in \fref{fig:cosmology_compare} into separate cases to examine the constraining power from the velocity dispersion only (VD only), time-delay measurements only (TD only), and a joint constraint from both measurements (VD+TD). \fref{fig:cosmology_LCDM} clearly shows that the velocity dispersion constrains $\lambdaint$ when assuming a cosmological model (e.g., $\Lambda$CDM). 
In other words, the value of $\lambdaint$ depends on the cosmological model, and the measurement of $\Ddt$ in this case is not a cosmological-model-independent quantity.
%The most recent analysis from TDCOSMO uses the $\Lambda$CDM model to break the internal MST and to obtain a determination of $H_{0}$ \citep{BirrerEtal20}.

% \subsection{Cosmologically-independent $\Dd$ and $\Ddt$ measurements}
\subsection{Method 2: Using external data sets to constrain $\DsDds$}
%To take into account possibly statistical non-zero $\lambdaint$ and simultaneously 
To break the internal MST without assuming a particular cosmological model (e.g., $\Lambda$CDM model), we require additional information to constrain $\DsDds$.  
%in addition to velocity dispersion, extra information which 
%In \sref{sec:kappaint_cosmo_dependent}, we show that a cosmological model provide the constraint on $\DsDds$ and hence $\lambdaint$ is constrained by velocity dispersion. 
This can be done by including data on Type Ia supernovae (SN1a) and Baryon acoustic oscillations (BAO). 
%on Type Ia supernovae (SNe Ia). 
%The idea of this method is similar to \citet{ArendseEtal19} and \citet{TaubenbergerEtal19}, where
% they use distance measurements from TDC to anchor the absolute distance and hence measure $H_{0}$. 
%they use inverse distance ladder approach, using strong-lensing time-delay distance measurements to constrain the normalization of the relative distance indicators such as SN1a data and hence constrain $H_{0}$. 
%The major difference under the presence of internal MST is that these external data sets are affecting the determination of $\lambdaint$ and also the time-delay distance.
%Without the calibration provided by Cepheids or Tip of the Red Giant Branch (TRGB) stars, SNe Ia constrain the shape of the expansion history (i.e., $H(z)/H_{0}$) rather than the normalization \citep[i.e., $H_0$;][]{CuestaEtal15}.
%We can use this information to set the value of $\DsDds$.
%\citet{BetouleEtal14} provide a compilation of 740 Type Ia Supernovae, including 239 supernovae from the SuperNovae Legacy Survey (SNLS), 374 from the SDSS, 118 supernovae from low-redshift surveys, and some beyond $z>1$ from HST. These are binned in 31 bins equally spaced in log(1 + z) as in the appendix of \citet{BetouleEtal14}.
The SN1a data are given as measurements
of the distance modulus
\begin{equation}
    \mu(z)\equiv m-M=25+5{\rm log}_{10}D_{L}(z),
\end{equation}
where $m$ is the apparent magnitude, $M$ is a fiducial absolute magnitude, and $D_{L}$ is the luminosity distance. 
When $M$ is a free parameter without calibration, SN1a only constrain the shape of the expansion history.
The BAO data provide measurements of $D_{V}/r_{\rm s}$, the dilation scale normalized by the standard ruler length (or $D_{\rm A}/r_{\rm s}$ and $Hr_{\rm s}$ in the anisotropic analysis), where
\begin{equation}
    D_{V}\equiv \left[ D_{\rm M}^{2}(z)\frac{cz}{H(z)}\right]^{1/3},
\end{equation}
$D_{\rm M}=(1+z)D_{\rm A}$, and $D_{\rm A}$ is the angular diameter distance.
If we vary $M$ and $r_{\rm s}$ freely, those data sets provide the information on the shape of the expansion history \citep[e.g.,][]{CuestaEtal15} and thus $\DsDds$.

However, we do still require some model for the redshift distance
relationship to connect these data.
In this work, we choose piece-wise natural cubic splines\footnote{Note that linear interpolation \citep[e.g.][]{VerdeEtal17}, Gaussian Processes \citep[e.g.][]{JoudakiEtal18,LiaoEtal20}, or smooth Taylor expansion \citep[e.g.][]{MacaulayEtal19,WojtakAgnello19,ArendseEtal19} are alternatives.} to describe $H(z)$ that fit to the data. 
The spline method has been used in many studies to reconstruct the expansion history \citep{BernalEtal16,PoulinEtal18,AylorEtal19,BernalEtal19}.
The splines are set by the values they take at different redshifts. These values can uniquely define the piece-wise cubic spline once we require continuity of $H(z)$ and its first and second derivatives at the knots, and set two boundary conditions.  
We also require the second derivative to vanish at the exterior knots. 
We set the minimal assumptions of this work: 
\begin{enumerate}
    \item Cosmological principle of homogeneity and isotropy (i.e., Friedmann–Lemaître–Robertson–Walker metric),
    \item Assumption of general relativity: the curvature density parameter is given by $\Omega_{k}=k[c/(R_{0}H_{0})]^{2}$  where $k=\{1,0,-1\}$, $R_{0}$ denotes the present value of the scale factor,
    \item Spline $H(z)$ completely specifies the FLRW metric.
    \item Cosmic distance duality relation: $D_{L}=D_{A}(1+z)^{2}$.
\end{enumerate}

To get a good constraint on the cosmological-model-independent $\DsDds$ at the redshift of the mock lens, we need the data that cover the redshift up to the source redshift\footnote{For the current 7 TDCOSMO lenses, the source redshifts are all below $z_{\rm s}=2.5$.} ($z_{\rm s}=1.523$ in this case). Current existing data show that one can constrain the shape of the expansion history up to $\sim z=2.5$ \citep[see Fig. 3 in][]{BernalEtal19}.
Therefore, we update the likelihood used in \citet{BernalEtal19} and use
Pantheon data sets of SN1a \citep{ScolnicEtal18}; BAO from galaxies \citep{KazinEtal14,AlamEtal17,Gil-MarnEtal20}, quasars \citep{HouEtal20,NeveuxEtal20} and the Lyman-$\alpha$ forest \citep{duMasdesBourbouxEtal20}. 
For the eBOSS likelihoods, we use the Gaussian approximation and that BAO can be used to apply cosmological models beyond LCDM \citep{BernalEtal20,CarterEtal20}.
We summarize the BAO measurements and the redshift information in \tref{tab:BAO}. 
We set 5 “knots” at different redshifts ($z_{0}=0$, $z_{1}=0.25$, $z_{2}=0.5$, $z_{3}=1.$, $z_{4}=2.5$). 
The complete set of parameters for the Spline model is
$\{H_{0},H_{1},H_{2},H_{3},H_{4}, r_{\rm s},\Omega_{k},M\}$. 
Uniform priors are assumed for all parameters. 
We show the reconstructed $E(z) (=H(z)/H_{0})$ normalised by the values for our fiducial cosmology given by the best-fitting parameters from the Planck analysis for a $\Lambda$CDM model in \fref{fig:cubic_JLA}. The posterior of $\DsDds$ can be obtained by integrating $E(z)$.

\begin{table*}
\centering

\begin{tabular}{lrc} 
\hline
 \hline
 Measurement & $z_{\rm eff}$ &  Reference\\
 \hline
 $r_{\rm s}/D_{V}$ & $0.106$ & \citet{BeutlerEtal11} \\ 
 \hline
 $D_{V}/r_{\rm s}$ & $0.15$ & \citet{RossEtal15} \\ 
 \hline
 $D_{M}(r_{\rm s,fid}/r_{\rm s})$ (Mpc) & $0.38$ & \citet{AlamEtal17} \\ 
 $H(z)(r_{\rm s}/r_{\rm s, fid})$ ($\kmsmpc$) & $0.38$ & \\
 $D_{M}(r_{\rm s,fid}/r_{\rm s})$ (Mpc) & $0.51$ &  \\ 
 $H(z)(r_{\rm s}/r_{\rm s, fid})$ ($\kmsmpc$) & $0.51$ &  \\ 
 $D_{M}(r_{\rm s,fid}/r_{\rm s})$ (Mpc) & $0.61$ & \\ 
 $H(z)(r_{\rm s}/r_{\rm s, fid})$ ($\kmsmpc$) & $0.61$ & \\ 
 \hline
 $D_{V}/r_{\rm s}$ & $0.44$ &  \citet{KazinEtal14} \\
 $D_{V}/r_{\rm s}$ & $0.6$ &   \\
 $D_{V}/r_{\rm s}$ & $0.73$ &   \\
 \hline
 $D_{M}/r_{\rm s}$ & $0.698$ &  \citet{Gil-MarnEtal20}  \\
 $c/(H(z)*r_{\rm s})$ & $0.698$ & \\
 \hline
 $D_{M}/r_{\rm s}$ & $1.48$  & \citet{HouEtal20} and \citet{NeveuxEtal20}  \\
 $c/(H(z)*r_{\rm s})$ & $1.48$ & \\
 \hline
 $D_{M}/r_{\rm s}$ & $2.33$ &  \citet{duMasdesBourbouxEtal20}  \\
 $c/(H(z)*r_{\rm s})$ & $2.33$ &  \\
 \hline
 \hline
\end{tabular}
\caption{Summary of the BAO measurements that are used in this work. In our fiducial cosmology, $r_{{\rm s}, {\rm fid}} = 147.78$ Mpc.}
\label{tab:BAO}
\end{table*}

%We then use compressed form of the JLA likelihood to reconstruct $H(z)/H_{0}$ shown in \fref{fig:cubic_JLA}. 

In \fref{fig:Dd_JLA_Ddtprior}, we show that by combining the inference from the external datasets on $H(z)/H_{0}$, which constrain $\DsDds$, with time-delay strong lensing systems, one can obtain cosmological-model-independent $\Dd$ and $\Ddt$ measurements that are comparable to those obtained by assuming a $\Lambda$CDM or $w$CDM cosmology.
In addition, we also see that the values of $\lambdaint$ under $\Lambda$CDM and $w$CDM are slightly offset from the cases which include SN1a and BAO data sets. This is because the flat priors on $\Omega_{\rm m}$ and $w$ in $\Lambda$CDM and $w$CDM models do not reflect the expansion history described by the SN1a and BAO datasets. Thus, it indicates the importance of including the external datasets which directly constrain the expansion history to get distance measurements.

In \fref{fig:hz_recon}, we show the distance measurements from combining a single time-delay lens with SN1a and BAO with free $\Omega_{k}$ (the green contour in \fref{fig:Dd_JLA_Ddtprior}). This distance measurements can be used to infer $H_{0}$ in generic cosmological models. 
We emphasize that this approach does not require the absolute calibration of SN1a or BAO; thus, the derived constraint on $H_0$ remains independent of the distance ladder and the sound horizon scale.

\subsection{Comparison with Paper IV}
In the previous section, we demonstrate that $\DsDds$ is fully degenerate with $\lambdaint$, which affects the time-delay distance measurement. 
Therefore, we compare the redshift-dependent median value of $\DsDds$, constrained directly by the SN and BAO data at the redshift of the current 7 TDCOSMO lens samples, with the $\DsDds$ in Paper IV, which used the prior based on Pantheon sample with $\Omega_{\rm m}=0.298\pm0.022$ \citep{ScolnicEtal18} under the assumption of $\Lambda$CDM model. 
These results shown in \fref{fig:deviation_LCDM_omegaPantheon} demonstrate that in the case of these 7 lenses, using the prior information from Pantheon data sets which constrain the low redshift expansion history, and then using $\Lambda$CDM model to extrapolate the constraint on $\DsDds$ to high redshift are valid approaches since the deviations from the median value of $\DsDds$ are all below 1\% demonstrating that the shape of the expansion history is well-described by the $\Lambda$CDM model.
However, the time-delay distance measurements derived by the method developed in this work are broadly applicable distance posteriors, which can be used to infer $H_{0}$ in various cosmological models.

\section{Conclusions}
\label{sec:conclusion}
In this work, we use a mock gravitational lens system to study the correlation between distance measurements under the mass-sheet transformation (MST) with or without assuming a cosmological model.
We verify that although $\Ddt$ is directly correlated with both the internal and external MST, $\Dd$ is not only invariant under an external MST but is also insensitive to the internal MST. 
Thus, without assuming any particular cosmological model, the role of velocity dispersion is to obtain the angular diameter distance to the lens ($\Dd$) rather than break the internal MST ($\lambdaint$). 
To break $\lambdaint$, in addition to the velocity dispersion, we identify that constraining $\DsDds$ is the key, which is typically achieved through the assumption of a particular cosmological model, and hence $\lambdaint$ and $\Ddt$ are both cosmological-model-dependent quantities. 
In this work, we show that cosmological-model-independent $\Ddt$ measurement can be achieved when one uses relative distance indicators (e.g., SN1a and BAO) to constrain $\DsDds$ and hence $\lambdaint$.
These distance measurements with SN1a and BAO shown in \fref{fig:hz_recon} can then be used to infer $H_{0}$ in generic cosmological models. 
It is important to stress that this approach does not require the absolute calibration of SN1a or BAO; thus, the derived constraint on $H_0$ remains independent of the distance ladder and the sound horizon scale.

\begin{acknowledgements}
GC-FC thanks James Chan, Tommaso Treu, Dominique Sluse, Matt Auger, and Elizabeth Buckley-Geer for many insightful comments and espeically Simon Birrer and Adriano Agnello for  
GC-FC thanks Alessandro Sonnenfeld for useful suggestions on using stellar kinematics code\footnote{\url{https://github.com/astrosonnen/spherical_jeans}} and Ken Wong for the code review. GC-FC acknowledges support by NSF through grants NSF-AST-1906976 and NSF-AST-1836016. GC-FC acknowledges support from the Moore Foundation through grant 8548.
GC-FC and CDF acknowledge support for this work from the National Science Foundation under Grant Nos. AST-171561 and AST-1907396. 
SHS and AY thank the Max Planck Society for support through the Max Planck Research Group for SHS. 
SHS and EK are supported in part by the Deutsche Forschungsgemeinschaft (DFG, German Research Foundation) under Germany's Excellence Strategy - EXC-2094 - 390783311. 
The Kavli IPMU is supported by World Premier International Research Center Initiative (WPI), MEXT, Japan. 
JLB is supported by the Allan C.\ and Dorothy H.\ Davis Fellowship.
\end{acknowledgements}

% WARNING
%-------------------------------------------------------------------
% Please note that we have included the references to the file aa.dem in
% order to compile it, but we ask you to:
%
% - use BibTeX with the regular commands:
\bibliographystyle{aa} % style aa.bst
\bibliography{AO_cosmography} % your references Yourfile.bib

\begin{thebibliography}{75}
\expandafter\ifx\csname natexlab\endcsname\relax\def\natexlab#1{#1}\fi

\bibitem[{{Alam} {et~al.}(2017){Alam}, {Ata}, {Bailey}, {Beutler}, {Bizyaev},
  {Blazek}, {Bolton}, {Brownstein}, {Burden}, {Chuang}, {Comparat}, {Cuesta},
  {Dawson}, {Eisenstein}, {Escoffier}, {Gil-Mar{\'\i}n}, {Grieb}, {Hand}, {Ho},
  {Kinemuchi}, {Kirkby}, {Kitaura}, {Malanushenko}, {Malanushenko}, {Maraston},
  {McBride}, {Nichol}, {Olmstead}, {Oravetz}, {Padmanabhan},
  {Palanque-Delabrouille}, {Pan}, {Pellejero-Ibanez}, {Percival}, {Petitjean},
  {Prada}, {Price- Whelan}, {Reid}, {Rodr{\'\i}guez- Torres}, {Roe}, {Ross},
  {Ross}, {Rossi}, {Rubi{\~n}o-Mart{\'\i}n}, {Saito}, {Salazar-Albornoz},
  {Samushia}, {S{\'a}nchez}, {Satpathy}, {Schlegel}, {Schneider},
  {Sc{\'o}ccola}, {Seo}, {Sheldon}, {Simmons}, {Slosar}, {Strauss}, {Swanson},
  {Thomas}, {Tinker}, {Tojeiro}, {Maga{\~n}a}, {Vazquez}, {Verde}, {Wake},
  {Wang}, {Weinberg}, {White}, {Wood-Vasey}, {Y{\`e}che}, {Zehavi}, {Zhai}, \&
  {Zhao}}]{AlamEtal17}
{Alam}, S., {Ata}, M., {Bailey}, S., {et~al.} 2017, \mnras, 470, 2617

\bibitem[{{Arendse} {et~al.}(2019){Arendse}, {Wojtak}, {Agnello}, {Chen},
  {Fassnacht}, {Sluse}, {Hilbert}, {Millon}, {Bonvin}, {Wong}, {Courbin},
  {Suyu}, {Birrer}, {Treu}, \& {Koopmans}}]{ArendseEtal19}
{Arendse}, N., {Wojtak}, R.~J., {Agnello}, A., {et~al.} 2019, arXiv e-prints,
  arXiv:1909.07986

\bibitem[{{Auger} {et~al.}(2010){Auger}, {Treu}, {Bolton}, {Gavazzi},
  {Koopmans}, {Marshall}, {Moustakas}, \& {Burles}}]{AugerEtal10}
{Auger}, M.~W., {Treu}, T., {Bolton}, A.~S., {et~al.} 2010, \apj, 724, 511

\bibitem[{{Aylor} {et~al.}(2019){Aylor}, {Joy}, {Knox}, {Millea},
  {Raghunathan}, \& {Kimmy Wu}}]{AylorEtal19}
{Aylor}, K., {Joy}, M., {Knox}, L., {et~al.} 2019, \apj, 874, 4

\bibitem[{{Barkana}(1998)}]{Barkana98}
{Barkana}, R. 1998, \apj, 502, 531

\bibitem[{{Barnab{\`e}} {et~al.}(2011){Barnab{\`e}}, {Czoske}, {Koopmans},
  {Treu}, \& {Bolton}}]{BarnabeEtal11}
{Barnab{\`e}}, M., {Czoske}, O., {Koopmans}, L.~V.~E., {Treu}, T., \& {Bolton},
  A.~S. 2011, \mnras, 415, 2215

\bibitem[{{Bernal} {et~al.}(2019){Bernal}, {Breysse}, \&
  {Kovetz}}]{BernalEtal19}
{Bernal}, J.~L., {Breysse}, P.~C., \& {Kovetz}, E.~D. 2019, \prl, 123, 251301

\bibitem[{{Bernal} {et~al.}(2020){Bernal}, {Smith}, {Boddy}, \&
  {Kamionkowski}}]{BernalEtal20}
{Bernal}, J.~L., {Smith}, T.~L., {Boddy}, K.~K., \& {Kamionkowski}, M. 2020,
  arXiv e-prints, arXiv:2004.07263

\bibitem[{{Bernal} {et~al.}(2016){Bernal}, {Verde}, \& {Riess}}]{BernalEtal16}
{Bernal}, J.~L., {Verde}, L., \& {Riess}, A.~G. 2016, \jcap, 10, 019

\bibitem[{{Beutler} {et~al.}(2011){Beutler}, {Blake}, {Colless}, {Jones},
  {Staveley-Smith}, {Campbell}, {Parker}, {Saunders}, \&
  {Watson}}]{BeutlerEtal11}
{Beutler}, F., {Blake}, C., {Colless}, M., {et~al.} 2011, \mnras, 416, 3017

\bibitem[{{Birrer} {et~al.}(2020){Birrer}, {Shajib}, {Galan}, {Millon}, {Treu},
  {Agnello}, {Auger}, {Chen}, {Christensen}, {Collett}, {Courbin}, {Fassnacht},
  {Koopmans}, {Marshall}, {Park}, {Rusu}, {Sluse}, {Spiniello}, {Suyu},
  {Wagner-Carena}, {Wong}, {Barnab{\`e}}, {Bolton}, {Czoske}, {Ding},
  {Frieman}, \& {Van de Vyvere}}]{BirrerEtal20}
{Birrer}, S., {Shajib}, A.~J., {Galan}, A., {et~al.} 2020, arXiv e-prints,
  arXiv:2007.02941

\bibitem[{{Birrer} {et~al.}(2019){Birrer}, {Treu}, {Rusu}, {Bonvin},
  {Fassnacht}, {Chan}, {Agnello}, {Shajib}, {Chen}, {Auger}, {Courbin},
  {Hilbert}, {Sluse}, {Suyu}, {Wong}, {Marshall}, {Lemaux}, \&
  {Meylan}}]{BirrerEtal19}
{Birrer}, S., {Treu}, T., {Rusu}, C.~E., {et~al.} 2019, \mnras, 484, 4726

\bibitem[{{Blum} {et~al.}(2020){Blum}, {Castorina}, \&
  {Simonovi{\'c}}}]{BlumEtal20}
{Blum}, K., {Castorina}, E., \& {Simonovi{\'c}}, M. 2020, \apjl, 892, L27

\bibitem[{{Bolton} {et~al.}(2006){Bolton}, {Burles}, {Koopmans}, {Treu}, \&
  {Moustakas}}]{BoltonEtal06SDSS}
{Bolton}, A.~S., {Burles}, S., {Koopmans}, L. V.~E., {Treu}, T., \&
  {Moustakas}, L.~A. 2006, \apj, 638, 703

\bibitem[{{Bolton} {et~al.}(2004){Bolton}, {Burles}, {Schlegel}, {Eisenstein},
  \& {Brinkmann}}]{BoltonEtal04SDSS}
{Bolton}, A.~S., {Burles}, S., {Schlegel}, D.~J., {Eisenstein}, D.~J., \&
  {Brinkmann}, J. 2004, \aj, 127, 1860

\bibitem[{{Bonvin} {et~al.}(2018){Bonvin}, {Chan}, {Millon}, {Rojas},
  {Courbin}, {Chen}, {Fassnacht}, {Paic}, {Tewes}, {Chao}, {Chijani}, {Gilman},
  {Gilmore}, {Williams}, {Buckley-Geer}, {Frieman}, {Marshall}, {Suyu}, {Treu},
  {Hempel}, {Kim}, {Lachaume}, {Rabus}, {Anguita}, {Meylan}, {Motta}, \&
  {Magain}}]{BonvinEtal18_PGTD}
{Bonvin}, V., {Chan}, J.~H.~H., {Millon}, M., {et~al.} 2018, \aap, 616, A183

\bibitem[{{Bonvin} {et~al.}(2016){Bonvin}, {Tewes}, {Courbin}, {Kuntzer},
  {Sluse}, \& {Meylan}}]{BonvinEtal16}
{Bonvin}, V., {Tewes}, M., {Courbin}, F., {et~al.} 2016, \aap, 585, A88

\bibitem[{{Buckley-Geer} {et~al.}(2020){Buckley-Geer}, {Lin}, {Rusu}, {Poh},
  {Palmese}, {Agnello}, {Christensen}, {Frieman}, {Shajib}, {Treu}, {Collett},
  {Birrer}, {Anguita}, {Fassnacht}, {Meylan}, {Mukherjee}, {Wong}, {Aguena},
  {Allam}, {Avila}, {Bertin}, {Bhargava}, {Brooks}, {Carnero Rosell}, {Carrasco
  Kind}, {Carretero}, {Castander}, {Costanzi}, {da Costa}, {De Vicente},
  {Desai}, {Diehl}, {Doel}, {Eifler}, {Everett}, {Flaugher}, {Fosalba},
  {Garc{\'\i}a-Bellido}, {Gaztanaga}, {Gruen}, {Gruendl}, {Gschwend},
  {Gutierrez}, {Hinton}, {Honscheid}, {James}, {Kuehn}, {Kuropatkin}, {Maia},
  {Marshall}, {Melchior}, {Menanteau}, {Miquel}, {Ogand o}, {Paz-Chinch{\'o}n},
  {Plazas}, {Sanchez}, {Scarpine}, {Schubnell}, {Serrano}, {Sevilla-Noarbe},
  {Smith}, {Soares-Santos}, {Suchyta}, {Swanson}, {Tarle}, {Tucker}, {Varga},
  \& {DES Collaboration}}]{Buckley-GeerEtal20}
{Buckley-Geer}, E.~J., {Lin}, H., {Rusu}, C.~E., {et~al.} 2020, \mnras, 498,
  3241

\bibitem[{{Cappellari}(2016)}]{Cappellari16}
{Cappellari}, M. 2016, \araa, 54, 597

\bibitem[{{Carter} {et~al.}(2020){Carter}, {Beutler}, {Percival}, {DeRose},
  {Wechsler}, \& {Zhao}}]{CarterEtal20}
{Carter}, P., {Beutler}, F., {Percival}, W.~J., {et~al.} 2020, \mnras, 494,
  2076

\bibitem[{{Chen} {et~al.}(2019){Chen}, {Fassnacht}, {Suyu}, {Rusu}, {Chan},
  {Wong}, {Auger}, {Hilbert}, {Bonvin}, {Birrer}, {Millon}, {Koopmans},
  {Lagattuta}, {McKean}, {Vegetti}, {Courbin}, {Ding}, {Halkola}, {Jee},
  {Shajib}, {Sluse}, {Sonnenfeld}, \& {Treu}}]{GChenEtal19}
{Chen}, G. C.~F., {Fassnacht}, C.~D., {Suyu}, S.~H., {et~al.} 2019, \mnras,
  2193

\bibitem[{{Collett} {et~al.}(2013){Collett}, {Marshall}, {Auger}, {Hilbert},
  {Suyu}, {Greene}, {Treu}, {Fassnacht}, {Koopmans}, {Brada{\v c}}, \&
  {Blandford}}]{CollettEtal13}
{Collett}, T.~E., {Marshall}, P.~J., {Auger}, M.~W., {et~al.} 2013, \mnras,
  432, 679

\bibitem[{{Cuesta} {et~al.}(2015){Cuesta}, {Verde}, {Riess}, \&
  {Jimenez}}]{CuestaEtal15}
{Cuesta}, A.~J., {Verde}, L., {Riess}, A., \& {Jimenez}, R. 2015, \mnras, 448,
  3463

\bibitem[{{du Mas des Bourboux} {et~al.}(2020){du Mas des Bourboux}, {Rich},
  {Font-Ribera}, {de Sainte Agathe}, {Farr}, {Etourneau}, {Le Goff}, {Cuceu},
  {Balland}, {Bautista}, {Blomqvist}, {Brinkmann}, {Brownstein}, {Chabanier},
  {Chaussidon}, {Dawson}, {Gonz{\'a}lez-Morales}, {Guy}, {Lyke}, {de la
  Macorra}, {Mueller}, {Myers}, {Nitschelm}, {Mu{\~n}oz Guti{\'e}rrez},
  {Palanque-Delabrouille}, {Parker}, {Percival}, {P{\'e}rez-R{\`a}fols},
  {Petitjean}, {Pieri}, {Ravoux}, {Rossi}, {Schneider}, {Seo}, {Slosar},
  {Stermer}, {Vivek}, {Y{\`e}che}, \& {Youles}}]{duMasdesBourbouxEtal20}
{du Mas des Bourboux}, H., {Rich}, J., {Font-Ribera}, A., {et~al.} 2020, \apj,
  901, 153

\bibitem[{{Efstathiou}(2020)}]{Efstathiou20}
{Efstathiou}, G. 2020, arXiv e-prints, arXiv:2007.10716

\bibitem[{{Falco} {et~al.}(1985){Falco}, {Gorenstein}, \&
  {Shapiro}}]{FalcoEtal85}
{Falco}, E.~E., {Gorenstein}, M.~V., \& {Shapiro}, I.~I. 1985, \apjl, 289, L1

\bibitem[{{Fassnacht} {et~al.}(2006){Fassnacht}, {Gal}, {Lubin}, {McKean},
  {Squires}, \& {Readhead}}]{FassnachtEtal06}
{Fassnacht}, C.~D., {Gal}, R.~R., {Lubin}, L.~M., {et~al.} 2006, \apj, 642, 30

\bibitem[{{Fassnacht} {et~al.}(2002){Fassnacht}, {Xanthopoulos}, {Koopmans}, \&
  {Rusin}}]{FassnachtEtal02}
{Fassnacht}, C.~D., {Xanthopoulos}, E., {Koopmans}, L.~V.~E., \& {Rusin}, D.
  2002, \apj, 581, 823

\bibitem[{{Freedman} {et~al.}(2019){Freedman}, {Madore}, {Hatt}, {Hoyt},
  {Jang}, {Beaton}, {Burns}, {Lee}, {Monson}, {Neeley}, {Phillips}, {Rich}, \&
  {Seibert}}]{FreedmanEtal19}
{Freedman}, W.~L., {Madore}, B.~F., {Hatt}, D., {et~al.} 2019, \apj, 882, 34

\bibitem[{{Gil-Mar{\'\i}n} {et~al.}(2020){Gil-Mar{\'\i}n}, {Bautista},
  {Paviot}, {Vargas-Maga{\~n}a}, {de la Torre}, {Fromenteau}, {Alam},
  {{\'A}vila}, {Burtin}, {Chuang}, {Dawson}, {Hou}, {de Mattia}, {Mohammad},
  {M{\"u}ller}, {Nadathur}, {Neveux}, {Percival}, {Raichoor}, {Rezaie}, {Ross},
  {Rossi}, {Ruhlmann-Kleider}, {Smith}, {Tamone}, {Tinker}, {Tojeiro}, {Wang},
  {Zhao}, {Zhao}, {Brinkmann}, {Brownstein}, {Choi}, {Escoffier}, {de la
  Macorra}, {Moon}, {Newman}, {Schneider}, {Seo}, \& {Vivek}}]{Gil-MarnEtal20}
{Gil-Mar{\'\i}n}, H., {Bautista}, J.~E., {Paviot}, R., {et~al.} 2020, \mnras,
  498, 2492

\bibitem[{{Gorenstein} {et~al.}(1988){Gorenstein}, {Falco}, \&
  {Shapiro}}]{GorensteinEtal88}
{Gorenstein}, M.~V., {Falco}, E.~E., \& {Shapiro}, I.~I. 1988, \apj, 327, 693

\bibitem[{{Greene} {et~al.}(2013){Greene}, {Suyu}, {Treu}, {Hilbert}, {Auger},
  {Collett}, {Marshall}, {Fassnacht}, {Blandford}, {Brada{\v c}}, \&
  {Koopmans}}]{GreeneEtal13}
{Greene}, Z.~S., {Suyu}, S.~H., {Treu}, T., {et~al.} 2013, \apj, 768, 39

\bibitem[{{Grillo} {et~al.}(2008){Grillo}, {Lombardi}, \&
  {Bertin}}]{GrilloEtal08}
{Grillo}, C., {Lombardi}, M., \& {Bertin}, G. 2008, \aap, 477, 397

\bibitem[{{Hinshaw} {et~al.}(2013){Hinshaw}, {Larson}, {Komatsu}, {Spergel},
  {Bennett}, {Dunkley}, {Nolta}, {Halpern}, {Hill}, {Odegard}, {Page}, {Smith},
  {Weiland}, {Gold}, {Jarosik}, {Kogut}, {Limon}, {Meyer}, {Tucker}, {Wollack},
  \& {Wright}}]{HinshawEtal13}
{Hinshaw}, G., {Larson}, D., {Komatsu}, E., {et~al.} 2013, \apjs, 208, 19

\bibitem[{{Hou} {et~al.}(2020){Hou}, {S{\'a}nchez}, {Ross}, {Smith}, {Neveux},
  {Bautista}, {Burtin}, {Zhao}, {Scoccimarro}, {Dawson}, {de Mattia}, {de la
  Macorra}, {du Mas des Bourboux}, {Eisenstein}, {Gil-Mar{\'\i}n}, {Lyke},
  {Mohammad}, {Mueller}, {Percival}, {Vargas Maga{\~n}a}, {Rossi}, {Zarrouk},
  {Zhao}, {Brinkmann}, {Brownstein}, {Chuang}, {Myers}, {Newman}, {Schneider},
  \& {Vivek}}]{HouEtal20}
{Hou}, J., {S{\'a}nchez}, A.~G., {Ross}, A.~J., {et~al.} 2020, arXiv e-prints,
  arXiv:2007.08998

\bibitem[{{Humphrey} \& {Buote}(2010)}]{HumphreyBuote10}
{Humphrey}, P.~J. \& {Buote}, D.~A. 2010, \mnras, 403, 2143

\bibitem[{{Jee} {et~al.}(2015){Jee}, {Komatsu}, \& {Suyu}}]{JeeEtal15}
{Jee}, I., {Komatsu}, E., \& {Suyu}, S.~H. 2015, \jcap, 11, 033

\bibitem[{{Jee} {et~al.}(2019){Jee}, {Suyu}, {Komatsu}, {Fassnacht}, {Hilbert},
  \& {Koopmans}}]{JeeEtal19}
{Jee}, I., {Suyu}, S.~H., {Komatsu}, E., {et~al.} 2019, Science, 365, 1134

\bibitem[{{Joudaki} {et~al.}(2018){Joudaki}, {Kaplinghat}, {Keeley}, \&
  {Kirkby}}]{JoudakiEtal18}
{Joudaki}, S., {Kaplinghat}, M., {Keeley}, R., \& {Kirkby}, D. 2018, \prd, 97,
  123501

\bibitem[{{Kazin} {et~al.}(2014){Kazin}, {Koda}, {Blake}, {Padmanabhan},
  {Brough}, {Colless}, {Contreras}, {Couch}, {Croom}, {Croton}, {Davis},
  {Drinkwater}, {Forster}, {Gilbank}, {Gladders}, {Glazebrook}, {Jelliffe},
  {Jurek}, {Li}, {Madore}, {Martin}, {Pimbblet}, {Poole}, {Pracy}, {Sharp},
  {Wisnioski}, {Woods}, {Wyder}, \& {Yee}}]{KazinEtal14}
{Kazin}, E.~A., {Koda}, J., {Blake}, C., {et~al.} 2014, \mnras, 441, 3524

\bibitem[{{Komatsu} {et~al.}(2011){Komatsu}, {Smith}, {Dunkley}, {Bennett},
  {Gold}, {Hinshaw}, {Jarosik}, {Larson}, \& {et al.}}]{KomatsuEtal11}
{Komatsu}, E., {Smith}, K.~M., {Dunkley}, J., {et~al.} 2011, \apjs, 192, 18

\bibitem[{{Koopmans} {et~al.}(2009){Koopmans}, {Bolton}, {Treu}, {Czoske},
  {Auger}, {Barnab{\`e}}, {Vegetti}, {Gavazzi}, \& {et al.}}]{KoopmansEtal09}
{Koopmans}, L.~V.~E., {Bolton}, A., {Treu}, T., {et~al.} 2009, \apjl, 703, L51

\bibitem[{{Koopmans} {et~al.}(2006){Koopmans}, {Treu}, {Bolton}, {Burles}, \&
  {Moustakas}}]{KoopmansEtal06}
{Koopmans}, L.~V.~E., {Treu}, T., {Bolton}, A.~S., {Burles}, S., \&
  {Moustakas}, L.~A. 2006, \apj, 649, 599

\bibitem[{{Liao} {et~al.}(2020){Liao}, {Shafieloo}, {Keeley}, \&
  {Linder}}]{LiaoEtal20}
{Liao}, K., {Shafieloo}, A., {Keeley}, R.~E., \& {Linder}, E.~V. 2020, \apjl,
  895, L29

\bibitem[{{Macaulay} {et~al.}(2019){Macaulay}, {Nichol}, {Bacon}, {Brout},
  {Davis}, {Zhang}, {Bassett}, {Scolnic}, {M{\"o}ller}, {D'Andrea}, {Hinton},
  {Kessler}, {Kim}, {Lasker}, {Lidman}, {Sako}, {Smith}, {Sullivan}, {Abbott},
  {Allam}, {Annis}, {Asorey}, {Avila}, {Bechtol}, {Brooks}, {Brown}, {Burke},
  {Calcino}, {Carnero Rosell}, {Carollo}, {Carrasco Kind}, {Carretero},
  {Castander}, {Collett}, {Crocce}, {Cunha}, {da Costa}, {Davis}, {De Vicente},
  {Diehl}, {Doel}, {Drlica-Wagner}, {Eifler}, {Estrada}, {Evrard},
  {Filippenko}, {Finley}, {Flaugher}, {Foley}, {Fosalba}, {Frieman}, {Galbany},
  {Garc{\'\i}a-Bellido}, {Gaztanaga}, {Glazebrook}, {Gonz{\'a}lez-Gait{\'a}n},
  {Gruen}, {Gruendl}, {Gschwend}, {Gutierrez}, {Hartley}, {Hollowood},
  {Honscheid}, {Hoormann}, {Hoyle}, {Huterer}, {Jain}, {James}, {Jeltema},
  {Kasai}, {Krause}, {Kuehn}, {Kuropatkin}, {Lahav}, {Lewis}, {Li}, {Lima},
  {Lin}, {Maia}, {Marshall}, {Martini}, {Miquel}, {Nugent}, {Palmese}, {Pan},
  {Plazas}, {Romer}, {Roodman}, {Sanchez}, {Scarpine}, {Schindler},
  {Schubnell}, {Serrano}, {Sevilla-Noarbe}, {Sharp}, {Soares-Santos},
  {Sobreira}, {Sommer}, {Suchyta}, {Swann}, {Swanson}, {Tarle}, {Thomas},
  {Thomas}, {Tucker}, {Uddin}, {Vikram}, {Walker}, \&
  {Wiseman}}]{MacaulayEtal19}
{Macaulay}, E., {Nichol}, R.~C., {Bacon}, D., {et~al.} 2019, \mnras, 966

\bibitem[{{Merritt}(1985)}]{Merritt85}
{Merritt}, D. 1985, \aj, 90, 1027

\bibitem[{{Millon} {et~al.}(2020){Millon}, {Galan}, {Courbin}, {Treu}, {Suyu},
  {Ding}, {Birrer}, {Chen}, {Shajib}, {Sluse}, {Wong}, {Agnello}, {Auger},
  {Buckley-Geer}, {Chan}, {Collett}, {Fassnacht}, {Hilbert}, {Koopmans},
  {Motta}, {Mukherjee}, {Rusu}, {Sonnenfeld}, {Spiniello}, \& {Van de
  Vyvere}}]{MillonEtal20}
{Millon}, M., {Galan}, A., {Courbin}, F., {et~al.} 2020, \aap, 639, A101

\bibitem[{{Navarro} {et~al.}(1996){Navarro}, {Frenk}, \&
  {White}}]{NavarroEtal96}
{Navarro}, J.~F., {Frenk}, C.~S., \& {White}, S.~D.~M. 1996, \apj, 462, 563

\bibitem[{{Neveux} {et~al.}(2020){Neveux}, {Burtin}, {de Mattia}, {Smith},
  {Ross}, {Hou}, {Bautista}, {Brinkmann}, {Chuang}, {Dawson}, {Gil-Mar{\'\i}n},
  {Lyke}, {de la Macorra}, {du Mas des Bourboux}, {Mohammad}, {M{\"u}ller},
  {Myers}, {Newman}, {Percival}, {Rossi}, {Schneider}, {Vivek}, {Zarrouk},
  {Zhao}, \& {Zhao}}]{NeveuxEtal20}
{Neveux}, R., {Burtin}, E., {de Mattia}, A., {et~al.} 2020, \mnras, 499, 210

\bibitem[{{Osipkov}(1979)}]{Osipkov79}
{Osipkov}, L.~P. 1979, Pis ma Astronomicheskii Zhurnal, 5, 77

\bibitem[{{Planck Collaboration} {et~al.}(2018){Planck Collaboration},
  {Aghanim}, {Akrami}, {Ashdown}, {Aumont}, {Baccigalupi}, {Ballardini},
  {Banday}, {Barreiro}, {Bartolo}, {Basak}, {Battye}, {Benabed}, {Bernard},
  {Bersanelli}, {Bielewicz}, {Bock}, {Bond}, {Borrill}, {Bouchet}, {Boulanger},
  {Bucher}, {Burigana}, {Butler}, {Calabrese}, {Cardoso}, {Carron},
  {Challinor}, {Chiang}, {Chluba}, {Colombo}, {Combet}, {Contreras}, {Crill},
  {Cuttaia}, {de Bernardis}, {de Zotti}, {Delabrouille}, {Delouis}, {Di
  Valentino}, {Diego}, {Dor{\'e}}, {Douspis}, {Ducout}, {Dupac}, {Dusini},
  {Efstathiou}, {Elsner}, {En{\ss}lin}, {Eriksen}, {Fantaye}, {Farhang},
  {Fergusson}, {Fernandez-Cobos}, {Finelli}, {Forastieri}, {Frailis},
  {Franceschi}, {Frolov}, {Galeotta}, {Galli}, {Ganga}, {G{\'e}nova-Santos},
  {Gerbino}, {Ghosh}, {Gonz{\'a}lez-Nuevo}, {G{\'o}rski}, {Gratton},
  {Gruppuso}, {Gudmundsson}, {Hamann}, {Handley}, {Herranz}, {Hivon}, {Huang},
  {Jaffe}, {Jones}, {Karakci}, {Keih{\"a}nen}, {Keskitalo}, {Kiiveri}, {Kim},
  {Kisner}, {Knox}, {Krachmalnicoff}, {Kunz}, {Kurki-Suonio}, {Lagache},
  {Lamarre}, {Lasenby}, {Lattanzi}, {Lawrence}, {Le Jeune}, {Lemos},
  {Lesgourgues}, {Levrier}, {Lewis}, {Liguori}, {Lilje}, {Lilley}, {Lindholm},
  {L{\'o}pez-Caniego}, {Lubin}, {Ma}, {Mac{\'\i}as-P{\'e}rez}, {Maggio},
  {Maino}, {Mandolesi}, {Mangilli}, {Marcos-Caballero}, {Maris}, {Martin},
  {Martinelli}, {Mart{\'\i}nez- Gonz{\'a}lez}, {Matarrese}, {Mauri}, {McEwen},
  {Meinhold}, {Melchiorri}, {Mennella}, {Migliaccio}, {Millea}, {Mitra},
  {Miville-Desch{\^e}nes}, {Molinari}, {Montier}, {Morgante}, {Moss}, {Natoli},
  {N{\o}rgaard-Nielsen}, {Pagano}, {Paoletti}, {Partridge}, {Patanchon},
  {Peiris}, {Perrotta}, {Pettorino}, {Piacentini}, {Polastri}, {Polenta},
  {Puget}, {Rachen}, {Reinecke}, {Remazeilles}, {Renzi}, {Rocha}, {Rosset},
  {Roudier}, {Rubi{\~n}o-Mart{\'\i}n}, {Ruiz-Granados}, {Salvati}, {Sandri},
  {Savelainen}, {Scott}, {Shellard}, {Sirignano}, {Sirri}, {Spencer},
  {Sunyaev}, {Suur-Uski}, {Tauber}, {Tavagnacco}, {Tenti}, {Toffolatti},
  {Tomasi}, {Trombetti}, {Valenziano}, {Valiviita}, {Van Tent}, {Vibert},
  {Vielva}, {Villa}, {Vittorio}, {Wandelt}, {Wehus}, {White}, {White},
  {Zacchei}, \& {Zonca}}]{planck18parameter}
{Planck Collaboration}, {Aghanim}, N., {Akrami}, Y., {et~al.} 2018, arXiv
  e-prints, arXiv:1807.06209

\bibitem[{{Poulin} {et~al.}(2018){Poulin}, {Boddy}, {Bird}, \&
  {Kamionkowski}}]{PoulinEtal18}
{Poulin}, V., {Boddy}, K.~K., {Bird}, S., \& {Kamionkowski}, M. 2018, \prd, 97,
  123504

\bibitem[{{Refsdal}(1964)}]{Refsdal64}
{Refsdal}, S. 1964, \mnras, 128, 307

\bibitem[{{Riess} {et~al.}(2019){Riess}, {Casertano}, {Yuan}, {Macri}, \&
  {Scolnic}}]{RiessEtal19}
{Riess}, A.~G., {Casertano}, S., {Yuan}, W., {Macri}, L.~M., \& {Scolnic}, D.
  2019, arXiv e-prints, arXiv:1903.07603

\bibitem[{{Ross} {et~al.}(2015){Ross}, {Samushia}, {Howlett}, {Percival},
  {Burden}, \& {Manera}}]{RossEtal15}
{Ross}, A.~J., {Samushia}, L., {Howlett}, C., {et~al.} 2015, \mnras, 449, 835

\bibitem[{{Rusu} {et~al.}(2017){Rusu}, {Fassnacht}, {Sluse}, {Hilbert}, {Wong},
  {Huang}, {Suyu}, {Collett}, {Marshall}, {Treu}, \& {Koopmans}}]{RusuEtal17}
{Rusu}, C.~E., {Fassnacht}, C.~D., {Sluse}, D., {et~al.} 2017, \mnras, 467,
  4220

\bibitem[{{Rusu} {et~al.}(2019){Rusu}, {Wong}, {Bonvin}, {Sluse}, {Suyu},
  {Fassnacht}, {Chan}, {Hilbert}, {Auger}, {Sonnenfeld}, {Birrer}, {Courbin},
  {Treu}, {Chen}, {Halkola}, {Koopmans}, {Marshall}, \&
  {Shajib}}]{RusuEtal19_H0LiCOW}
{Rusu}, C.~E., {Wong}, K.~C., {Bonvin}, V., {et~al.} 2019, \mnras, 498, 1440

\bibitem[{{Schneider} \& {Sluse}(2013)}]{SchneiderSluse13}
{Schneider}, P. \& {Sluse}, D. 2013, \aap, 559, A37

\bibitem[{{Scolnic} {et~al.}(2018){Scolnic}, {Jones}, {Rest}, {Pan},
  {Chornock}, {Foley}, {Huber}, {Kessler}, {Narayan}, {Riess}, {Rodney},
  {Berger}, {Brout}, {Challis}, {Drout}, {Finkbeiner}, {Lunnan}, {Kirshner},
  {Sand ers}, {Schlafly}, {Smartt}, {Stubbs}, {Tonry}, {Wood-Vasey}, {Foley},
  {Hand}, {Johnson}, {Burgett}, {Chambers}, {Draper}, {Hodapp}, {Kaiser},
  {Kudritzki}, {Magnier}, {Metcalfe}, {Bresolin}, {Gall}, {Kotak}, {McCrum}, \&
  {Smith}}]{ScolnicEtal18}
{Scolnic}, D.~M., {Jones}, D.~O., {Rest}, A., {et~al.} 2018, \apj, 859, 101

\bibitem[{{Shajib} {et~al.}(2020{\natexlab{a}}){Shajib}, {Birrer}, {Treu},
  {Agnello}, {Buckley-Geer}, {Chan}, {Christensen}, {Lemon}, {Lin}, {Millon},
  {Poh}, {Rusu}, {Sluse}, {Spiniello}, {Chen}, {Collett}, {Courbin},
  {Fassnacht}, {Frieman}, {Galan}, {Gilman}, {More}, {Anguita}, {Auger},
  {Bonvin}, {McMahon}, {Meylan}, {Wong}, {Abbott}, {Annis}, {Avila}, {Bechtol},
  {Brooks}, {Brout}, {Burke}, {Carnero Rosell}, {Carrasco Kind}, {Carretero},
  {Castander}, {Costanzi}, {da Costa}, {De Vicente}, {Desai}, {Dietrich},
  {Doel}, {Drlica-Wagner}, {Evrard}, {Finley}, {Flaugher}, {Fosalba},
  {Garc{\'\i}a-Bellido}, {Gerdes}, {Gruen}, {Gruendl}, {Gschwend}, {Gutierrez},
  {Hollowood}, {Honscheid}, {Huterer}, {James}, {Jeltema}, {Krause},
  {Kuropatkin}, {Li}, {Lima}, {MacCrann}, {Maia}, {Marshall}, {Melchior},
  {Miquel}, {Ogando}, {Palmese}, {Paz-Chinch{\'o}n}, {Plazas}, {Romer},
  {Roodman}, {Sako}, {Sanchez}, {Santiago}, {Scarpine}, {Schubnell}, {Scolnic},
  {Serrano}, {Sevilla-Noarbe}, {Smith}, {Soares-Santos}, {Suchyta}, {Tarle},
  {Thomas}, {Walker}, \& {Zhang}}]{ShajibEtal20_0408}
{Shajib}, A.~J., {Birrer}, S., {Treu}, T., {et~al.} 2020{\natexlab{a}}, \mnras,
  494, 6072

\bibitem[{{Shajib} {et~al.}(2020{\natexlab{b}}){Shajib}, {Treu}, {Birrer}, \&
  {Sonnenfeld}}]{ShajibEtal20}
{Shajib}, A.~J., {Treu}, T., {Birrer}, S., \& {Sonnenfeld}, A.
  2020{\natexlab{b}}, arXiv e-prints, arXiv:2008.11724

\bibitem[{{Sonnenfeld} {et~al.}(2013){Sonnenfeld}, {Treu}, {Gavazzi}, {Suyu},
  {Marshall}, {Auger}, \& {Nipoti}}]{SonnenfeldEtal13}
{Sonnenfeld}, A., {Treu}, T., {Gavazzi}, R., {et~al.} 2013, \apj, 777, 98

\bibitem[{{Suyu} {et~al.}(2013){Suyu}, {Auger}, {Hilbert}, {Marshall}, {Tewes},
  {Treu}, {Fassnacht}, {Koopmans}, {Sluse}, {Blandford}, {Courbin}, \&
  {Meylan}}]{SuyuEtal13}
{Suyu}, S.~H., {Auger}, M.~W., {Hilbert}, S., {et~al.} 2013, \apj, 766, 70

\bibitem[{{Suyu} {et~al.}(2018){Suyu}, {Chang}, {Courbin}, \&
  {Okumura}}]{SuyuEtal18}
{Suyu}, S.~H., {Chang}, T.-C., {Courbin}, F., \& {Okumura}, T. 2018, \ssr, 214,
  91

\bibitem[{{Suyu} {et~al.}(2010){Suyu}, {Marshall}, {Auger}, {Hilbert},
  {Blandford}, {Koopmans}, {Fassnacht}, \& {Treu}}]{SuyuEtal10}
{Suyu}, S.~H., {Marshall}, P.~J., {Auger}, M.~W., {et~al.} 2010, \apj, 711, 201

\bibitem[{{Suyu} {et~al.}(2009){Suyu}, {Marshall}, {Blandford}, {Fassnacht},
  {Koopmans}, {McKean}, \& {Treu}}]{SuyuEtal09}
{Suyu}, S.~H., {Marshall}, P.~J., {Blandford}, R.~D., {et~al.} 2009, \apj, 691,
  277

\bibitem[{{Suyu} {et~al.}(2014){Suyu}, {Treu}, {Hilbert}, {Sonnenfeld},
  {Auger}, {Blandford}, {Collett}, {Courbin}, {Fassnacht}, {Koopmans},
  {Marshall}, {Meylan}, {Spiniello}, \& {Tewes}}]{SuyuEtal14}
{Suyu}, S.~H., {Treu}, T., {Hilbert}, S., {et~al.} 2014, \apjl, 788, L35

\bibitem[{{Taubenberger} {et~al.}(2019){Taubenberger}, {Suyu}, {Komatsu},
  {Jee}, {Birrer}, {Bonvin}, {Courbin}, {Rusu}, {Shajib}, \&
  {Wong}}]{TaubenbergerEtal19}
{Taubenberger}, S., {Suyu}, S.~H., {Komatsu}, E., {et~al.} 2019, \aap, 628, L7

\bibitem[{{Tihhonova} {et~al.}(2018){Tihhonova}, {Courbin}, {Harvey},
  {Hilbert}, {Rusu}, {Fassnacht}, {Bonvin}, {Marshall}, {Meylan}, {Sluse},
  {Suyu}, {Treu}, \& {Wong}}]{TihhonovaEtal18}
{Tihhonova}, O., {Courbin}, F., {Harvey}, D., {et~al.} 2018, \mnras, 477, 5657

\bibitem[{{Treu} \& {Marshall}(2016)}]{TreuMarshall16}
{Treu}, T. \& {Marshall}, P.~J. 2016, \aapr, 24, 11

\bibitem[{{Verde} {et~al.}(2017){Verde}, {Bernal}, {Heavens}, \&
  {Jimenez}}]{VerdeEtal17}
{Verde}, L., {Bernal}, J.~L., {Heavens}, A.~F., \& {Jimenez}, R. 2017, \mnras,
  467, 731

\bibitem[{{Verde} {et~al.}(2019){Verde}, {Treu}, \& {Riess}}]{VerdeEtal19}
{Verde}, L., {Treu}, T., \& {Riess}, A.~G. 2019, Nature Astronomy, 3, 891

\bibitem[{{Wojtak} \& {Agnello}(2019)}]{WojtakAgnello19}
{Wojtak}, R. \& {Agnello}, A. 2019, \mnras, 486, 5046

\bibitem[{Wong {et~al.}(2019)Wong, Suyu, Chen, Rusu, Millon, Sluse, Bonvin,
  Fassnacht, Taubenberger, Auger, Birrer, Chan, Courbin, Hilbert, Tihhonova,
  Treu, Agnello, Ding, Jee, Komatsu, Shajib, Sonnenfeld, Blandford, Koopmans,
  Marshall, \& Meylan}]{WongEtal19}
Wong, K.~C., Suyu, S.~H., Chen, G. C.-F., {et~al.} 2019, Monthly Notices of the
  Royal Astronomical Society, 498, 1420

\bibitem[{{Xu} {et~al.}(2016){Xu}, {Sluse}, {Schneider}, {Springel},
  {Vogelsberger}, {Nelson}, \& {Hernquist}}]{XuEtal16}
{Xu}, D., {Sluse}, D., {Schneider}, P., {et~al.} 2016, \mnras, 456, 739

\end{thebibliography}
%
% - join the .bib files when you upload your source files
%-------------------------------------------------------------------

\end{document}